\documentclass[acmsmall]{acmart}

\usepackage[utf8]{inputenc}
\usepackage{physics}
\usepackage{natbib} %for citaion

\usepackage {url}
\usepackage{pifont}
\usepackage{color} 
\usepackage{booktabs}
\usepackage{colortbl}
\usepackage{makecell}
\usepackage{multirow}
\usepackage{graphicx}

\usepackage{array}

\usepackage[figuresright]{rotating}

\usepackage{lscape}

\usepackage{afterpage}

\usepackage{makecell}

\usepackage{tocvsec2}

\usepackage{xcolor}

\definecolor{maroon}{cmyk}{0,0.87,0.68,0.32}

\acmJournal{JACM}
\acmYear{2023}
\setcopyright{acmcopyright}
\begin{document}

\title{Quantum Computing for Databases: Overview and Challenges}

\author{Gongsheng Yuan}
\orcid{0000-0002-7992-3179}
\affiliation{
  \institution{Zhejiang University}
  % \streetaddress{38 Zheda Road}
  \city{Hangzhou}
  \country{China}
}
\email{ygs@zju.edu.cn}

\author{Yuxing Chen}
\orcid{0000-0002-6220-2535}
\affiliation{
  \institution{Tencent Inc.}
  \streetaddress{Tower B Jindi Center, No. 16 High Tech South 10 Road}
  \city{Shenzhen}
  \country{China}
}
\email{axingguchen@tencent.com}

\author{Jiaheng Lu}
\authornote{Corresponding author}
\orcid{0000-0003-2067-454X}
\affiliation{
  \institution{University of Helsinki}
  \streetaddress{C212, Exactum, Kumpula, Gustaf Hallstromin katu 2b}
  \city{Helsinki}
  \country{Finland}
}
\email{jiaheng.lu@helsinki.fi}

\author{Sai Wu}
\orcid{0000-0002-7903-1496}
\email{wusai@zju.edu.cn}
\affiliation{%
  \institution{Zhejiang University}
  \city{Hangzhou}
  \country{China}
}

% \author{Chang Yao}
% \orcid{0000-0002-1187-6257}
% \email{changy@zju.edu.cn}
% \affiliation{%
%   \institution{Zhejiang University}
%   \city{Hangzhou}
%   \country{China}
%   % \postcode{310027}
% }

\author{Zhiwei Ye}
% \orcid{0000-0002-1187-6257}
\email{yezhiwei@cmss.chinamobile.com }
\affiliation{%
  \institution{China Mobile (Suzhou) Software Technology Co., Ltd}
  \city{Suzhou}
  \country{China}
  % \postcode{310027}
}

\author{Ling Qian}
% \orcid{0000-0002-1187-6257}
\email{qianling@cmss.chinamobile.com}
\affiliation{%
  \institution{China Mobile (Suzhou) Software Technology Co., Ltd}
  \city{Suzhou}
  \country{China}
  % \postcode{310027}
}

\author{Gang Chen}%
\orcid{0000-0002-7483-0045}
\email{cg@zju.edu.cn}
\affiliation{%
  \institution{Zhejiang University}
  \city{Hangzhou}
  \country{China}
  % \postcode{310027}
}

\renewcommand{\shortauthors}{G. Yuan et al.}

\begin{abstract}

In the decades, the general field of quantum computing has experienced remarkable progress since its inception. A plethora of researchers not only proposed quantum algorithms showing the power of quantum computing but also constructed the prototype of quantum computers, making it walk into our tangible reality. Those remarkable advancements in quantum computing have opened doors for novel applications, one of which is quantum databases. Researchers are trying to use a paradigm brought by quantum computing to revolutionize various aspects of database management systems. In this paper, we envision the synergy between quantum computing and databases with two perspectives: Quantum computing-enabled technology, and quantum computing-inspired technology. Based on this classification, we present a detailed overview of the research attained in this area, aiming to show the landscape of the field and draw a road map of future directions.

% The general field of quantum computing has experienced
% remarkable progress in the last years becoming a tangible
% reality. Prototypes of quantum computers, also known as
% noisy intermediate-scale quantum (NISQ) computers [1], already exist and have been made available to users through the cloud [2], [3]. We will still have to wait for having largescale and fault-tolerant quantum computers that provide the expected computational power, but the potential of this new
% technology is undeniable [4]–[6]. A quantum computer will
% not only be capable of solving relevant problems unsolvable
% by current classical computers, it also represents a paradigm
% shift in the way of how computing is performed. In this paper, we envision the synergy between quantum computing and databases with two perspectives: Quantum computing enabled technologies and quantum computing inspired approaches.
% This paper presents a survey of the research done in this area, aiming to show the landscape of the field and draw a road-map of future directions.

\end{abstract}



\date{ \today}


\maketitle

\section{Introduction}

%\textbf{Gongsheng: write the first paragraph to summarize the remarkable progress of quantum computing and quantum machines } 

In the early 1980s, Paul Anthony Benioff of Argonne National Laboratory in the United States proposed the first quantum mechanical computer model \cite{Benioff1980TheCA}, which established the groundwork for subsequent quantum computing research. Since then, the field of quantum computing has experienced remarkable progress in these decades. Distinguished researchers have not only proposed some quantum algorithms (e.g., Shor's Algorithm \cite{365700}), theoretically proving that quantum computers have the ability to solve certain problems more effectively than classical computers, but the prototype of quantum computers has also become a tangible reality, being available to access through the cloud.

\afterpage{
\begin{landscape}\centering

\begin{table}
\centering
\setlength{\extrarowheight}{0pt}
\addtolength{\extrarowheight}{\aboverulesep}
\addtolength{\extrarowheight}{\belowrulesep}
\setlength{\aboverulesep}{0pt}
\setlength{\belowrulesep}{0pt}
\caption{The Technical Routes of Quantum Computers}
\label{tab:quantumComputers}

\resizebox{\columnwidth}{!}{%

\begin{tabular}{m{6em}<{\centering}ccccccc}
\midrule
% \specialrule{0.05em}{5pt}{5pt}

% \rowcolor[gray]{.9} 

% \textbf{} & \textbf{\begin{tabular}[c]{@{}c@{}}Superconducting   \\ quantum computer\end{tabular}} & \textbf{\begin{tabular}[c]{@{}c@{}}Trapped Ion\\ quantum computer\end{tabular}} & \textbf{\begin{tabular}[c]{@{}c@{}}Photonic   \\ quantum computer\end{tabular}} & \textbf{\begin{tabular}[c]{@{}c@{}}Neutral-atom\\ quantum computer\end{tabular}} & \textbf{\begin{tabular}[c]{@{}c@{}}Semiconductor\\ quantum computer\end{tabular}} & \textbf{\begin{tabular}[c]{@{}c@{}}Topological\\ quantum computer\end{tabular}} & \textbf{\begin{tabular}[c]{@{}c@{}}\textcolor{gray}{Quantum Annealer} \\ \textcolor{gray}{Special computer}\end{tabular}} \\

\rowcolor[gray]{.9} 
\cellcolor[gray]{.9}  & \cellcolor[gray]{.9}  & \cellcolor[gray]{.9}  & \cellcolor[gray]{.9}  & \cellcolor[gray]{.9}  & \cellcolor[gray]{.9}  & \cellcolor[gray]{.9}  & \cellcolor[gray]{.9}  \\
\rowcolor[gray]{.9}  
\cellcolor[gray]{.9}  & \cellcolor[gray]{.9}  & \cellcolor[gray]{.9}  & \cellcolor[gray]{.9}  & \cellcolor[gray]{.9}  & \cellcolor[gray]{.9}  & \cellcolor[gray]{.9}  & \cellcolor[gray]{.9}  \\
\rowcolor[gray]{.9}
\multirow{-3}{*}{\cellcolor[gray]{.9} \textbf{}} & \multirow{-3}{*}{\cellcolor[gray]{.9} \textbf{\begin{tabular}[c]{@{}c@{}}Superconducting   \\ quantum computer\end{tabular}}} & \multirow{-3}{*}{\cellcolor[gray]{.9} \textbf{\begin{tabular}[c]{@{}c@{}}Trapped Ion\\ quantum computer\end{tabular}}} & \multirow{-3}{*}{\cellcolor[gray]{.9} \textbf{\begin{tabular}[c]{@{}c@{}}Photonic   \\ quantum computer\end{tabular}}} & \multirow{-3}{*}{\cellcolor[gray]{.9} \textbf{\begin{tabular}[c]{@{}c@{}}Neutral-atom\\ quantum computer\end{tabular}}} & \multirow{-3}{*}{\cellcolor[gray]{.9} \textbf{\begin{tabular}[c]{@{}c@{}}Semiconductor\\ quantum computer\end{tabular}}} & \multirow{-3}{*}{\cellcolor[gray]{.9} \textbf{\begin{tabular}[c]{@{}c@{}}Topological\\ quantum computer\end{tabular}}} & \multirow{-3}{*}{\cellcolor[gray]{.9} \textbf{\begin{tabular}[c]{@{}c@{}}Quantum Annealer  \\ Special computer\end{tabular}}} \\

% \toprule

\specialrule{0.1em}{0pt}{5pt}
% \specialrule{0.1em}{5pt}{5pt}

\textbf{Time} & 2022 & 2022 & 2023 & 2022 & 2022 & - & 2023 \\ \specialrule{0.05em}{5pt}{5pt}

\textbf{Qubits} & 433 & 32 & 255 & 324 & 12 & - & 5000+ \\ \specialrule{0.05em}{5pt}{5pt}

\textbf{Representatives} & IBM & IonQ & USTC & Pasqal & Intel & Microsoft & D-Wave \\ \specialrule{0.05em}{5pt}{5pt}
 
\textbf{Country} & USA & USA & China & France & USA & USA & Canada \\ \specialrule{0.05em}{5pt}{5pt}

\textbf{Advantages} & \begin{tabular}[c]{@{}c@{}}Great scalability,\\ Longer coherence\\ time, Gate speed.\end{tabular} & \begin{tabular}[c]{@{}c@{}}High Qubit\\ Performance,\\ Exceptionally long\\ coherence times,\\ High gate fidelity.\end{tabular} & \begin{tabular}[c]{@{}c@{}}Room temperature\\ computation,\\ Error correction\\ flexibility,\\ Telecom compatibility.\end{tabular} & \begin{tabular}[c]{@{}c@{}}Long coherence \\ time, High scala-\\ bility. No need\\ for bulky cryogenic\\ equipment.\end{tabular} & \begin{tabular}[c]{@{}c@{}}Compatible with\\ current mature\\ semiconductor\\ technologies.\end{tabular} & \begin{tabular}[c]{@{}c@{}}Strong resistance \\ to external\\ interference.\end{tabular} & \begin{tabular}[c]{@{}c@{}}Solving\\ combinatorial\\ optimization\\ problems \end{tabular} \\ \specialrule{0.05em}{5pt}{5pt}

\textbf{Disadvantages} & \begin{tabular}[c]{@{}c@{}}Error rates,\\ Ultralow\\ temperature.\end{tabular} & Low scalability. & \begin{tabular}[c]{@{}c@{}}Difficult to\\ perform multi-\\ qubit operations.\end{tabular} & \begin{tabular}[c]{@{}c@{}}Difficult to\\ control individual\\ atoms precisely.\end{tabular} & \begin{tabular}[c]{@{}c@{}}Short coherence\\ time, Low\\ scalability.\end{tabular} & \begin{tabular}[c]{@{}c@{}}Research on it\\ is still theoretical\\ and experimental.\end{tabular} & \begin{tabular}[c]{@{}c@{}} Limited\\ qubit connectivity.\end{tabular} \\ \specialrule{0.05em}{5pt}{5pt}

\textbf{\begin{tabular}[c]{@{}c@{}}Other\\ Organizations\\ (e.g.,)\end{tabular}} & \begin{tabular}[c]{@{}c@{}}Google\\ (USA)\\ USTC\\ (China) \\ SEEQC \\ (USA) \\ Rigetti \\ (USA) \\ IQM Quantum Computers \\ (Finland)\end{tabular} & \begin{tabular}[c]{@{}c@{}}Quantinuum\\ (USA) \\ eleQtron \\ (Germany) \\ Universal Quantum \\ (UK) \\ Alpine Quantum Technologies \\ (Austria) \end{tabular} & \begin{tabular}[c]{@{}c@{}}Xanadu\\ (Canada)\\ QuiX Quantum\\ (Netherlands) \\ PsiQuantum \\ (USA) \\ Quandela \\ (France) \end{tabular} & \begin{tabular}[c]{@{}c@{}}QuEra\\ (USA) \\ ColdQuanta \\ (USA) \\ Atom Computing \\ (USA) \end{tabular} & \begin{tabular}[c]{@{}c@{}}QuTech\\ (Netherlands)\\ Archer Materials\\ (Australia) \\ Hitachi \\ (Japan) \\ EeroQ \\ (USA) \end{tabular} & \begin{tabular}[c]{@{}c@{}}Paul Scherrer Institute\\ (Switzerland)\\ IOPCAS \\ (China)\\ University of Chicago\\ (USA)\end{tabular} & - \\
\specialrule{0.15em}{5pt}{5pt}

\end{tabular}%
}
\end{table}

\end{landscape}
}

In terms of quantum machines, there is still a long way to go. The current mainstream technical routes of quantum computers includes superconducting \cite{devoret2004superconducting, harris2010experimental}, trapped ion \cite{cirac1995quantum, bruzewicz2019trapped}, photonic \cite{o2007optical, o2009photonic, madsen2022quantum}, neutral-atom \cite{barnes2022assembly, bluvstein2022quantum, graham2022multi}, semiconductor \cite{de2021materials, petit2020universal, veldhorst2017silicon, zwerver2022qubits}, and topological quantum computers \cite{kitaev2003fault, lahtinen2017short}, among which superconducting quantum computers are developing most rapidly (see Table~\ref{tab:quantumComputers}). The latest advances in the manufacture of quantum computers always attract much attention in various fields. So far,

\begin{itemize}
    \item The \textit{D-Wave} Corporation \cite{DWave} in Canada holds the distinction of having launched the world's first commercial quantum computer, ``D-Wave One'', on May 11, 2011. In 2023, the corporation proceeded to introduce a quantum computer boasting over 5000 qubits. However, it is crucial to note that quantum annealers, such as those produced by D-Wave, are predominantly designed to address optimization problems.

    \item \textit{IBM} \cite{IBM} emerged as a trailblazer in the realm of quantum computing, rolling out the IBM Q Experience in 2016, the world's first quantum cloud platform. Presently, the organization possesses over 60 operational quantum computers, surpassing the sum of all other companies worldwide. IBM's quantum computing cloud service has been instrumental to more than 100 commercial and research entities. Furthermore, IBM recently unveiled its most advanced quantum computer processor, the 433-qubit ``Osprey''.

    \item On June 22, 2023, \textit{IonQ} \cite{IonQ}, recognized as the world's pioneer ion trap quantum computing company, launched the IonQ Forte system equipped with 32 qubits. This system can be accessed via Amazon Braket, Microsoft Azure, and Google Cloud, or through direct API access.

    \item \textit{The University of Science and Technology of China} (USTC) \cite{deng2023gaussian} has reported new developments in Gaussian boson sampling experiments with pseudo-photon-number-resolving detection, which have witnessed up to 255 photon-click events.

    \item Founded in 2019, \textit{PASQAL} \cite{PASQAL} specializes in the construction of quantum computers from ordered neutral atoms in 2D and 3D arrays. This approach is aimed at delivering a practical quantum advantage to its clients and addressing real-world challenges. The organization announced the launch of a 324-atom (qubit) quantum processor, which held the record for the world's largest quantum processor until November 2022 when it was surpassed by IBM's 433-qubit superconducting quantum computer chip.

    \item \textit{Intel} \cite{Intel}, a significant player in the semiconductor industry, has achieved a key breakthrough in chip production research. The yield rate of their semiconductor qubit chip has reached an impressive 95\%, setting a new record for the number of silicon spin qubits at 12, doubling the previous record of 6 qubits \cite{philips2022universal} (\textit{QuTech} \cite{QuTech}). This achievement brings silicon spin qubit chips closer to mass production, a critical advancement towards generating the thousands or potentially millions of qubits required for commercial quantum computers.

    \item \textit{Microsoft} \cite{Microsoft}, on the other hand, is exploring a different approach. The company is attempting to encode its qubits in a type of quasiparticle, specifically \textit{non-abelian anyons}. However, the existence of these quasiparticles is still a subject of debate among physicists. Nevertheless, Microsoft aims to leverage the topological properties of these particles, which provide quantum states with robust resistance to external interference, to construct topological quantum computers.
\end{itemize}

These approaches to quantum computer implementation are obtained from recently published proposals/papers and commercially available devices. These implementations fall into two broad categories based on their computational models - universal and non-universal quantum computers. Universal quantum computers (e.g., gate-based systems utilizing superconducting or trapped ion qubits) aim to simulate a quantum Turing machine efficiently \cite{deutsch1985quantum}. In contrast, non-universal models like quantum annealers are problem-specific and cannot efficiently perform such universal quantum simulations.

The significant noise and errors in quantum hardware have led many researchers to characterize the present era of emerging quantum computing as the Noisy Intermediate-Scale Quantum (NISQ) regime \cite{Preskill_2018}. During this developmental phase, quantum devices have qubit counts surpassing the threshold where quantum advantages for specific tasks can start to emerge, but coherence times are still too short. It is an exciting yet challenging time for the field as scientists and engineers work to optimize qubit control and suppression of environmental perturbations, with the goal of building higher fidelity quantum machines capable of outperforming classical computers for increasingly complex problems. Constructing large-scale, fault-tolerant quantum computers is a crucial step toward practical quantum computing applications. While we are making steady progress towards this objective, such advanced computers are not yet within our reach soon. However, the current trajectory of development in quantum computing augurs well for the future. The potential of quantum computing is truly transformative, capable of not only tackling problems that are currently intractable for classical computers but also fundamentally altering the way we perform computations.

\begin{table}[!tbp]
\centering
\setlength{\extrarowheight}{0pt}
\addtolength{\extrarowheight}{\aboverulesep}
\addtolength{\extrarowheight}{\belowrulesep}
\setlength{\aboverulesep}{0pt}
\setlength{\belowrulesep}{0pt}
\caption{An Overview of Current Development of Quantum Computing for Databases}
\label{tab:overview}
\begin{tabular}{m{12em}<{\centering}m{14em}<{\centering}m{10em}<{\centering}}
\midrule
% \rowcolor{lightgray!20}
\rowcolor[gray]{.9}
\textbf{Problems} & \textbf{Quantum-Enabled} & \textbf{Quantum-Inspired}
\\ \toprule

Database Search & \cite{DBLP:conf/stoc/Grover96,Grover_1997,Grover_1997Single_Query,Terhal_1998,Boyer_1998,patel2001quantum,DBLP:journals/computing/ImreB04,1032254,Ju2007QuantumCD} & - \\
\hline

Database Manipulation & \cite{9712025,arxiv.0705.4303,arxiv.0710.3301,Pang2008QuantumSA,Salman2012QuantumSI,Salman2012QuantumSI,Gueddana2010OptimizedMF, DBLP:conf/adbis/JoczikK20} &  -  \\ 
\hline

Query Optimization & \cite{DBLP:journals/pvldb/Trummer016,arxiv.2107.10508,DBLP:conf/sigmod/Schonberger22,sigmod2023, 10.1145/3579142.3594298, winker2023quantum, DBLP:conf/vldb/SchonbergerTM23,DBLP:conf/vldb/GruenwaldWCGG23} &  \cite{mohsin2020dynamic,mohsin2021dynamic}  \\ 
\hline

Database Security & \cite{DBLP:journals/compsys/Ozhigov97,reif2009quantum,DBLP:conf/icdis/NjorbuenwuSZ19} &  -  \\ 
\hline

Transaction Management & \cite{OJCC_2020v7i1n01_Bittner,Bittner2020,Groppe2021Grover} &  -  \\ 
\hline

Resource Allocation & - &  \cite{DBLP:conf/cidr/0002KK13}  \\ 
\hline

Keyword Searches & - &  \cite{yuan2020quantum,yuan2021quantum} \\ 
\hline

\hline

IR Model and User Interaction & - &  \cite{DBLP:books/daglib/0011947, 10.1145/2484028.2484098, DBLP:conf/aaaifs/WangSK10, DBLP:conf/ecir/WangWH018}  \\ 
\hline

\hline
\bottomrule
\end{tabular}
\end{table}

Quantum computing, indeed, has witnessed many significant achievements since its inception, promising to revolutionize various domains, such as cryptography \cite{boneh1995quantum, zhao2023demonstration}, optimization \cite{li2020quantum}, materials science \cite{bauer2020quantum}, and chemistry \cite{cao2019quantum}. Those remarkable advancements in quantum computing have opened doors for novel applications (of academic or commercial interest), one of which is applying quantum computing to the field of databases \cite{DBLP:conf/vldb/YuanLCWYYL023}, an exploration of \textit{quantum practicality} \cite{hoefler2023disentangling}. Using quantum computing technologies in databases has the potential to offer significant advantages over classical databases, such as exponential speed-up in query processing, saving memory space, and enhanced data security. Despite the potential benefits, the development of this research direction is still in its infancy, with a limited number of prototypes and proof-of-concept implementations. The primary challenges hindering its growth include limited coherence times, the lack of mature quantum computing hardware, and the absence of well-established query languages and data models tailored for quantum systems.

% To accelerate the development of quantum databases, researchers have been actively working on several fronts. They have been exploring novel approaches to implement quantum data structures and algorithms, seeking to improve the efficiency and scalability of quantum query processing. Besides, interdisciplinary collaborations between the fields of quantum computing and database management have been fostering innovative ideas and solutions. Integrating quantum computing into database management systems has the potential to significantly impact various aspects, including but not limited to quantum database search, manipulation, query optimization, security, and transaction management.

To accelerate the development of databases with quantum computing, researchers have been actively working on several fronts. They have been exploring novel approaches to implement quantum data structures and algorithms, seeking to improve the efficiency and scalability of quantum query processing. Besides, interdisciplinary collaborations between the fields of quantum computing and database management have been fostering various innovative ideas and solutions, including but not limited to quantum database search, database manipulation, query optimization, database security, and transaction management. The research topics given in Table~\ref{tab:overview} are all the research on databases that we know so far after reviewing the literature.

This paper investigates the potential impact of integrating quantum computing into database management systems and the applicability of quantum computing to various database problems. Specifically, there are currently two ways to combine the respective strengths of quantum computing and databases. On the one hand, the unique capabilities of quantum computation can be used to enhance existing and develop novel database techniques, called Quantum computing-enable techniques. On the other hand, the theory of quantum mechanics can be used to motivate new algorithms for databases, called Quantum computing-inspired techniques.

% The approaches applied in such databases are classified into two categories: quantum-enable and quantum-inspired techniques.
% Quantum-enable techniques can encode problems in databases into the input of quantum algorithms, allowing quantum methods to solve these problems. 
% Quantum-inspired techniques, on the other hand, improve the algorithm parts used in database optimization by incorporating quantum mechanisms.

\vspace{0.5em}

\textbf{\textit{Quantum computing-enabled technology}}

\begin{enumerate}
    \item \textbf{Quantum Database Search}: Quantum search algorithms like Grover's algorithm can provide a quadratic speedup in searching unsorted databases compared to their classical counterparts. This improvement can be highly beneficial for large-scale databases where search operations are frequent and computationally demanding.
    \item \textbf{Database Manipulation}: Quantum computing can also offer efficient manipulation operations such as insertion, deletion, and selection. Researchers have been exploring quantum algorithms and data structures for these operations, aiming to minimize the time complexity and resource requirements compared to classical approaches.
    \item \textbf{Database Query Optimization}: Quantum computing can be used to optimize complex database queries by exploiting quantum parallelism and entanglement. Quantum optimization algorithms, such as the quantum approximate optimization algorithm (QAOA), can potentially improve the performance of query execution plans and minimize resource utilization.
    \item \textbf{Database Security}: Quantum technologies can enhance database security through quantum encryption methods, such as quantum key distribution (QKD). These methods can provide provable security against eavesdropping, ensuring the confidentiality and integrity of sensitive data stored in databases.
    \item \textbf{Transaction Management}: Quantum computing can potentially improve transaction management by enabling efficient concurrency control and conflict resolution mechanisms. Quantum techniques, such as quantum walks, can be applied to model transaction dependencies and detect potential conflicts, thereby improving the overall performance and scalability of database systems.
\end{enumerate}

\vspace{0.5em}

\textbf{\textit{Quantum computing-inspired technology}}

\begin{enumerate}
    \item \textbf{Resource Allocation}: Quantum-inspired methods can help solve complex optimization problems that arise in resource allocation, such as memory, CPU, and storage, thereby improving the performance of database management systems.
    \item \textbf{Keyword Searches}: Quantum-inspired technologies like the quantum language model can be employed to model the queries and candidate answers and return the most related answers to users.
    \item \textbf{Query Optimization}: Quantum-inspired algorithms such as quantum-inspired ant colony algorithm can be used to optimize query execution plans, thereby improving the performance of database management systems.
\end{enumerate}

% %\textbf{Yuxing: discuss the main contribution of this survey paper, such as a taxonomy of quantum computing for databases, several challenges and research directions, etc}

% The main contributions of this survey paper are summarized as follows:
% \begin{itemize}
%     \item  Please clearly state three contributions here!
%      \item
%       \item
% \end{itemize}

A considerable amount of studies have focused on quantum computing as a solution to optimization problems in the field of databases as summarized in Table~\ref{tab:overview}. As an exemplification of a prevalent challenge in database systems, the task of determining specific properties of elements within exponentially large search spaces presents a significant optimization hurdle. Quantum computers (QCs) are believed to exhibit exceptional performance in such scenarios, including the optimization of database queries \cite{DBLP:journals/pvldb/Trummer016,arxiv.2107.10508,DBLP:conf/sigmod/Schonberger22,sigmod2023} and transaction scheduling \cite{OJCC_2020v7i1n01_Bittner,Bittner2020,Groppe2021Grover}. 
In light of this, the contribution of this survey is to comprehensively examine existing quantum computing approaches and their optimization techniques in databases, which can potentially address various challenges in database applications. 
Despite the fact that many methods have demonstrated some degree of applicability and effectiveness, we have also observed that further research is necessary before practical and scalable applications of quantum computing in databases can be established. For instance, the current hardware infrastructure is incapable of supporting large datasets, necessitating a greater number of qubits for large-scale computations \cite{sigmod2023}. Additionally, the current algorithm encoding methods are limited, thereby constraining the potential scope of applications. Consequently, the potential range of applications remains restricted.

\smallskip
\noindent \textbf{Main Contributions} ~~
This manuscript reviews the representative techniques of quantum computing in the field of databases. This comprehensive survey helps to inspire the development of new applications of quantum computing in the database domain, facilitate the creation of practical quantum applications, and serve as a technical reference for selecting and comparing existing database optimization techniques. The key contributions of this survey are as follows:

\begin{enumerate}
\item We introduce the optimization of quantum computing in the field of databases and provide a general taxonomy and key features of the relevant methods.
\item We discuss quantum computing-enabled and quantum computing-inspired technologies for accelerating database operations from various perspectives.
% \item We describe in detail the key features of existing representative methods that utilize quantum computing.
% \item We outline open research problems and demonstrate that quantum computing in the database domain remains a challenging research area with potential applications in various real-world use cases.
\item We have described open research problems and identified the challenging research directions for the future work.
\end{enumerate}

% \textcolor{red}{Yuxing: this paragraph can be deleted or heavily rewritten. It is not about the contribution of this survey paper.}  
% As an exemplification of a prevalent challenge in database systems, the task of determining specific properties of elements within exponentially large search spaces presents a significant optimization hurdle. Quantum computers (QCs) are believed to exhibit exceptional performance in such scenarios, including the optimization of database queries \cite{DBLP:journals/pvldb/Trummer016,arxiv.2107.10508,DBLP:conf/sigmod/Schonberger22,sigmod2023} and transaction scheduling \cite{OJCC_2020v7i1n01_Bittner,Bittner2020,Groppe2021Grover}. However, further research is imperative to establish practical and scalable applications of quantum computing in databases. For instance, the current hardware infrastructure lacks the capability to support large datasets, necessitating a greater number of qubits for large-scale computations \cite{sigmod2023}. Moreover, the existing algorithm encoding methods are constrained, thereby limiting the potential scope of applications.

%\textbf{Yuxing: discuss the related work, other survey for quantum computing, but no survey on database applications}

\smallskip

\noindent \textbf{Related work} ~~ During our research, we discovered a survey \cite{reif2009quantum} from a decade ago discussing the algorithms, technologies, and challenges associated with utilizing quantum computing for various applications such as information processing, cryptography, coding theory, and algorithms. Furthermore, a more recent survey by Njorbuenwu et al.~\cite{DBLP:conf/icdis/NjorbuenwuSZ19} examined quantum information security, which has proven to be a successful application within the field of quantum theory. We also found that quantum methods are becoming increasingly popular and relevant for database optimizations. A new tutorial has recently been presented on utilizing quantum computing to accelerate two specific database problems, i.e., query optimization and transaction scheduling \cite{ccalikyilmaz2023opportunities}. %Recently, a short survey and vision on quantum computing for databases \cite{qdsm23} have been published, providing a brief introduction to the most researched topics in this field. 
However, there is currently a lack of comprehensive surveys that thoroughly evaluate the extensive array of potential applications and challenges of quantum computing specifically from a database application perspective. The most relevant to us is a short survey and vision on quantum computing for databases \cite{DBLP:conf/vldb/YuanLCWYYL023}, which only provides a brief introduction to the most researched topics in this field.

\smallskip

\noindent \textbf{Organization of this paper} ~~ The rest of the paper is organized as follows.
Section \ref{sec:background} presents the background. 
Section \ref{sec:quantum-based} introduces the quantum-enabled methods applied in databases.
Section \ref{sec:quantum-inspired} introduces the quantum-inspired methods applied in databases.
Section \ref{sec:open-challenges_future-work} discusses the open challenges and future directions.
Section \ref{sec:conclusion} concludes the paper.

%Various machine learning approaches were proposed to accomplish that task, some of which are cited below. In this paper, to further extend these studies, we propose a quantum machine learning approach for join order optimization.

\section{Background}\label{sec:background}

Quantum computing is a fast-growing technology that is able to utilize the laws of quantum mechanics to solve certain types of complex problems much faster than classical supercomputers do, even some problems that are currently beyond the reach of classical supercomputers. Machines supporting quantum computations are known as quantum computers. But quantum computers are heavily dependent on quantum algorithms. Hence, the world of quantum computing contains hardware study and software development. And this hardware is very different from the classical computers we know well.

% Quantum computing is a rapidly-emerging technology that utilizes quantum mechanics to solve the complex problems that a classical supercomputer is not easy to solve. The research usually involves hardware research and application development. Quantum computing is able to solve certain types of complex problems much faster by harnessing the phenomena of quantum mechanics, such as superposition, interference, and entanglement. Devices that perform quantum computations are known as quantum computers, which nowadays are available (e.g., IBM quantum system) but are still under development. There already existed many successful speedup applications for machine learning (ML), optimization, and simulation of physical systems \cite{DBLP:conf/ijcnn/RamezaniSMRA20}. This paper, we focus to discuss the main optimization in database systems.

\subsection{Qubit}

The basic unit that quantum computers manipulate is a \textit{qubit} (quantum bit) instead of a \textit{bit} (binary digit) that classical computers work with. The qubit is conventionally denoted in the form $\ket{\psi} = \alpha \ket{0} + \beta \ket{1}$ -- a qubit state can be represented by a two-dimensional vector while physicists typically use \textit{Dirac} notation -- where the coefficients $\alpha$ and $\beta$ are complex numbers called \textit{probability amplitudes}. If $\alpha$ or $\beta$ is zero, effectively, we could regard the qubit as a classical bit; if both are nonzero, the qubit will be in a superposition state. When performing \textit{measurement} (observation) on a qubit, $\alpha \ket{0} + \beta \ket{1}$, the qubit state collapses to $\ket{0}$ with probability $\alpha^2$ or to $\ket{1}$ with probability $\beta^2$, which also shows the norm-squared correspondence between amplitudes and probabilities. And any valid qubit state satisfies $\alpha^2 +\beta^2 = 1$.

% A qubit system with $k$ qubits can be denoted in the form $\ket{q_{k-1}...q_1q_0}$, a complex-valued vector, which represents a superposition of $2^k$ states. It provides quantum parallelism (i.e., handling multiple states simultaneously) to quantum computers.

A qubit system with $k$ qubits can be denoted in the form 
$$ \ket{\psi} = \sum \limits_{q_{1},q_{2},...,q_{k}\in \{0,1\}} \alpha_{q_{1}q_{2}...q_{k}}\ket{q_1}_1 \otimes \ket{q_2}_2 \otimes \cdots \otimes \ket{q_k}_k, $$
where the basis vector of $i_{th}$ qubit corresponds to $\{\ket{0}_i, \ket{1}_i\}$, and $\ket{q_1}_1 \otimes \ket{q_2}_2 \otimes \cdots \otimes \ket{q_k}_k$, also written as $\ket{q_1q_2...q_k}$, represents a basis vector of multi-qubit space. Such a superposition state provides quantum parallelism (i.e., handling multiple states simultaneously) to quantum computers.

Based on quantum bits, several computation paradigms of quantum computing, like quantum circuit \cite{nielsen2001quantum}, adiabatic quantum computation \cite{albash2018adiabatic}, and quantum Turing machine \cite{benioff1982quantum}, etc., have been proposed, where the quantum circuit model is the most widely used model of quantum computation.

\begin{figure*}[!t]
\centering
\includegraphics[height=5cm]{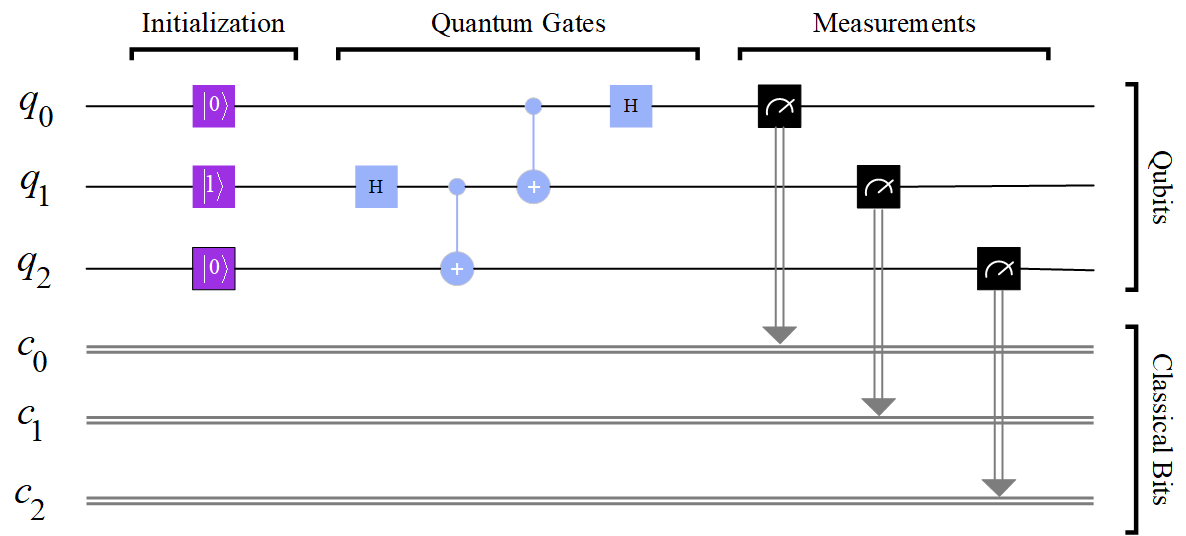}
\caption{An Example of Quantum Circuits}
\label{fig:quantumCircuit}
\end{figure*}

% quantum superposition, entanglement, interference.

\subsection{Quantum Circuit}

A qubit or a qubit system can alter its state via a series of unitary transformations. Since qubits are represented by column vectors mathematically, a unitary transformation is described by a matrix with complex entries. With this consensus, a quantum \textit{gate} will be defined as a unitary matrix to manipulate qubits. And a quantum computation can be represented as a network of quantum gates and measurements, i.e., a \textit{quantum circuit}.

A quantum circuit is a computational routine that executes the teleportation of qubits. In the circuit representation, it uses horizontal lines to represent qubits (double lines are classical bits), whose horizontal direction also indicates the flow of time when running a quantum algorithm, starting at the left-hand side of these lines and ending at the right. At the very beginning of each line, the initial state of the qubit is set. Then, selected gates or measurements are put on the corresponding qubits (lines) they act on, which should be accomplished in order (from left to right). Finally, the result of the algorithm is obtained at the right of the lines. Take a look at one example (Figure~\ref{fig:quantumCircuit}), the circuit representation \cite{nielsen2001quantum} is depicted by it well. 

In Figure~\ref{fig:quantumCircuit}, the quantum circuit has three qubits and three classical bits. Three main components exist in this circuit:

\begin{itemize}

    \item \textbf{Initialization} It initializes the qubit $q_i$ into a desired state by applying certain gates.
    
    \item\textbf{Quantum Gates} Based on a quantum algorithm, a sequence of quantum gates is put on corresponding qubits to manipulate these three qubits for achieving a specific goal. Here, it employs Hadamard gates, symbolized with an $H$, which function to put a qubit into superposition. Concurrently, it also utilizes CNOT gates, which either negate or leave the target qubit unchanged, as indicated by the $\oplus$ symbol, under the directive of the control qubit, designated by the $\cdot$ sign. This gate could create entangled states.

    \item \textbf{Measurements} To obtain the result of each qubit, it uses three classical bits to store the measurements of those three qubits as classical outcomes (0 or 1). Please note that it is the only non-unitary operation on a quantum circuit.
    
\end{itemize}

In this survey, most works focus on developing algorithms, building corresponding quantum circuits, and running them on quantum hardware or simulators.

\begin{figure*}[!t]
\centering
\includegraphics[height=8.2cm]{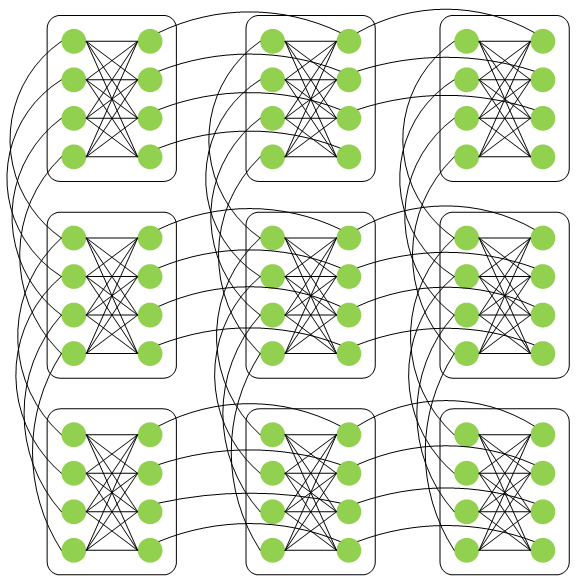}
\caption{An Extract of the Chimera Graph Structure}
\label{fig:quantumAnnealer}
\end{figure*}

\subsection{Quantum Annealer}

Adiabatic quantum computation (AQC) is another form of quantum computing, which encodes the solution of computational problems into the ground state of a time-dependent quantum Hamiltonian \cite{Vinci_2017}. It works based on the adiabatic theorem. However, implementing AQC using quantum physical systems is susceptible to non-ideal conditions that undermine the promise of the adiabatic theorem. Quantum annealing, enabling to capture a relaxation of the adiabatic condition, accommodates physical systems operating at finite temperatures and in open environments. This approach serves as a means to identify the minimum of an objective function, drawing upon the principles of Adiabatic Quantum Computation, without necessarily insisting on rigorous requirements (e.g., adiabaticity) \cite{AdiabaticQuantumComputingandQuantumAnnealing}.

% Adiabatic quantum computation is a form of quantum computing. It works based on the adiabatic theorem. In the framework of quantum mechanics, the idea of adiabaticity is inherently related to the idea of quantum annealing (an alternative to classical thermal simulated annealing).

% Quantum annealing can be compared to classical thermal simulated annealing. Classical thermal simulated annealing is a probabilistic technique that looks for the lowest point in the search space by putting the player at some point, letting the player find a new place based on local variations, and determining the player to move or freeze according to probability. However, such an algorithm may lead the player to a local minimum and needs numerous trials with different start points. 

% In contrast, quantum annealing starts with a superposition of all possible points (states) so that the amplitudes of all possible points (states) keep changing, realizing quantum parallelism. And quantum effects of entanglement and quantum tunneling could improve the outcome (e.g., quantum tunneling may make the player pass through hills if trapped in a local minimum).

% Quantum annealing starts with a superposition of all possible states so that the amplitudes of all possible points (states) keep changing, realizing quantum parallelism. And quantum fluctuations could bring the system out of the shallow local minima using tunneling.

Quantum annealing is an optimization process employed to address the challenge of finding the global minimum of a given objective function over a provided set of candidate solutions, commonly referred to as candidate states, using a process with quantum fluctuations. Quantum annealing is utilized primarily in scenarios where the problem's search space is discrete and with many local minima \cite{QAwikipedia}. A quantum annealer is a quantum device that employs the process of quantum annealing to tackle challenging optimization problems. The underlying methodology involves the evolution of a known initial configuration, performed at a non-zero temperature, towards the ground state of a Hamiltonian that encodes a specific problem \cite{boixo2013quantum}. D-Wave \cite{DWave}, a well-known implementation of quantum annealing. As shown in Firgure~\ref{fig:quantumAnnealer}, interactions between qubits in the D-Wave machine are limited to neighbors in a so-called \textit{Chimera} graph that could describe the qubit matrix of the quantum annealer. This structure is built on a unit cell. Each unit cell contains eight qubits. Those eight qubits form a complete bipartite graph. Besides, each left qubit of the unit cell is linked to the corresponding qubit in the above (and below) unit cell. In contrast, each right quilt of the unit cell is horizontally connected to the corresponding qubits in the left (and right) unit cells (some differences exist in the boundary qubits).

%complex computational challenges

Quantum computers are revolutionizing applications in a variety of fields by offering more efficient and expedited resolutions to complex computational challenges (e.g., Traveling Salesman Problem \cite{srinivasan2018efficient}, Subset Sum Problem \cite{bernstein2013quantum}, and Boolean Satisfiability Problem \cite{su2016quantum}). Just as the world of classical computing needs an algorithm ecosystem to found a broad community of application developers and users, the world of quantum computing does as well \cite{DWave}. In the upcoming subsection, we will acquaint first-time readers of quantum computing with some rudimentary quantum algorithms as preliminaries for accessing subsequent quantum works in the field of databases. For those who are already acquainted with Grover's algorithm, feel free to commence your expedition with section \ref{sec:quantum-based}.

%% Figure Graph
\begin{figure*}[!t]
\centering
\includegraphics[width=\linewidth]{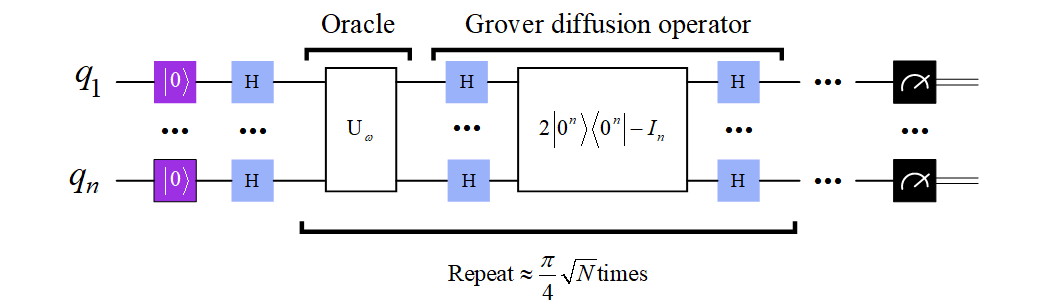}
\caption{Quantum circuit representation of Grover's algorithm}
\label{fig:groveralgorithm}
\end{figure*}

\subsection{Quantum Algorithm}

% A fast quantum mechanical algorithm for database search 1996
% \cite{DBLP:conf/stoc/Grover96}

\textit{\textbf{Grover's algorithm}}, also known as the quantum search algorithm, is proposed in \cite{DBLP:conf/stoc/Grover96, Grover_1997}, which could provide a quadratic speedup over the unsorted database. This is to say, given an unsorted database containing $N$ items out of which just one (e.g., $\omega$) is what we want to retrieve. Grover's algorithm could locate it in $O(\sqrt{N})$ quantum mechanical steps instead of checking the items in the database one by one $O(N)$.
To accomplish this acceleration, Grover's algorithm first initializes the system to the uniform superposition over all $N$ states where each state is represented as an $n$-bit string and only one state is what we want to find. Then it repeatedly performs the oracle operator and the Grover diffusion operator on those states around $\pi\sqrt{N}/4$ times to amplify the amplitude of the target state corresponding to the desired item. After that, it measures the resulting quantum state to obtain what we expect with a probability of at least $1/2$.

In this process, the oracle operator can be defined as:
\begin{equation}
    U_{\omega}\ket{x} = (-1)^{f(x)}\ket{x},
\end{equation}
where if $x$ = $\omega$, then $f(x) = 1$ and $U_{\omega}\ket{x} = -\ket{x}$; if $x$ $\neq$ $\omega$, then $f(x) = 0$ and $U_{\omega}\ket{x} = \ket{x}$. The Grover diffusion operator can be defined as:
\begin{equation}
    U_s = 2\ket{s}\bra{s} - I,
\end{equation}
where $\ket{s} = \tfrac{1}{\sqrt{N}}\sum_{x = 0}^{N - 1}\ket{x}$, that is, initial state, the uniform superposition over all $N$ states. This whole process is depicted in Figure~\ref{fig:groveralgorithm}.

% $\ket{s} = \tfrac{1}{\sqrt{N}}(\sum^{N - 1}\limits_{x = 0}\ket{x})$

\vspace{0.5em}

%Quadratic Unconstrained Binary Optimization problem

\textbf{\textit{Quadratic Unconstrained Binary Optimization} (QUBO)} \cite{lewis2017quadratic}, also referred to as unconstrained binary quadratic programming (UBQP), is a type of combinatorial optimization problem. It seeks to minimize a quadratic function of binary variables subject to no constraints, which can be represented as follows:
\begin{equation} 
    \mathop{\arg\min}\limits_{x \in \mathbb{B}}f_Q(x) = x^TQx = \sum\limits_{i = 1}^{n}\sum\limits_{j = i}^{n}Q_{ij}x_ix_j
\end{equation}
where $x$ is a binary vector of size $n$ whose elements are either 0 or 1 (i.e., $\mathbb{B} = \{0, 1\}$), and $Q$ is a symmetric matrix of size $n \times n$ with real values. The goal is to find the binary vector $x$ that minimizes the quadratic expression. 

QUBO problem is classified as NP-hard. Because of its close relationship to Ising models, QUBO constitutes a pivotal problem class for adiabatic quantum computation. It is solved using a physical process known as quantum annealing, which is a metaheuristic aiming to approximate the global optimum of a given function. In quantum annealing, the optimization problem is encoded in a quantum system, and the process seeks to find the system's lowest energy state (ground state), which corresponds to the optimal solution of the problem.

Quantum annealing starts in an arbitrary initial state. The transverse field Hamiltonian $H_{D}$ explores the optimization surface. As $H_{D}$ gradually guides the system towards its ground state, it remains there until it reaches the final Hamiltonian $H_{F}$, whose ground state corresponds to the optimal solution of the problem. The total Hamiltonian $H(t)$ of the system during the annealing process is given by:
\begin{equation}
    H(t) = H_{F} + \Gamma(t) H_{D}
\end{equation}
where the function $\Gamma(t)$ is the transverse field coefficient, which is initially very high and reduces to zero over time.

\vspace{0.5em}

%Hybrid algorithms run only partially on QPUs, and are augmented by CPU computations.

%the Quantum Approximate Optimization Algorithm

% \textit{\textbf{Quantum Approximate Optimization Algorithm}} 

\textbf{\textit{The Quantum Approximate Optimization Algorithm} (QAOA)} \cite{farhi2014quantum, zhou2020quantum} --- a hybrid quantum-classical variational algorithm --- is proposed to work with limited quantum resources and integrate them with classical routines for tackling combinatorial optimization problems (including QUBO problems). That is, QAOQ is a variational algorithm utilizing a \textit{parameterized quantum circuit} (i.e., Variational Quantum Circuits, VQC) \cite{benedetti2019parameterized}, which aims to find the best possible approximation to the optimal solution by adjusting the circuit's parameters and applying classical optimization techniques.

%That is, QAOQ uses the variational method to find the ground state of cost Hamiltonian that corresponds to the optimal solution of the optimization problem.

%% Figure Graph
\begin{figure*}[!t]
\centering
\includegraphics[width=0.95\linewidth]{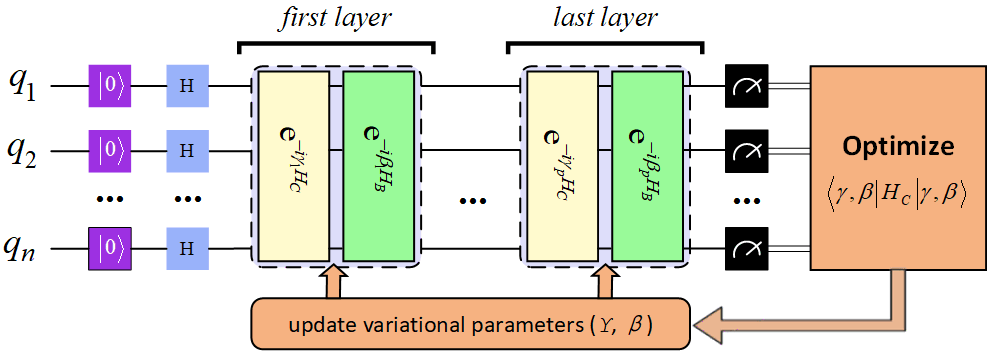}
\caption{An example of $p$-layer QAOQ}
\label{fig: QAOA}
\end{figure*}

% Specifically, QAOA involves two quantum Hamiltonians: one is the problem Hamiltonian $H_{C}$. Another is the mixer Hamiltonian $H_{B}$, required to be non-commuting with $H_{C}$. 

Specifically, QAOA involves two quantum Hamiltonians: 
\begin{itemize}
    \item One is the problem Hamiltonian $H_{C}$, encoding the classical objective function of the optimization problem as problem Hamiltonian.
    \item Another is the mixer Hamiltonian $H_{B}$. For example, it could be defined as the sum of Pauli operators X:
\begin{equation}
    H_B = \sum_{i = 1}^nX_i, \quad
    X = 
    \begin{bmatrix}
    0 & 1 \\
    1 & 0
\end{bmatrix}.
\end{equation}
\end{itemize}
 
Next, QAOA chooses any integer $p \ge 1$, prepares an initial state $\ket{+}^{\otimes n}$ (the uniform superposition over computational basis states, i.e., $\ket{+}^{\otimes n} = \tfrac{1}{\sqrt{2^n}}\sum_{x \in \{0, 1\}^n}\ket{x}$),
constructs the circuits $e^{-i\gamma H_C}$ (cost layer) and $e^{-i \beta H_B}$ (mixer layer) where $\gamma_1 \dots \gamma_p \equiv \gamma$ and $\beta_1 \dots \beta_p \equiv \beta$, and then defines parametric state $\ket{\gamma, \beta}$ as:
\begin{equation}
    \ket{\gamma, \beta} = \underbrace{e^{-i\beta_p H_B}e^{-i\gamma_p H_C}}_{last \ layer}\ \dots \ e^{-i\beta_i H_B}e^{-i\gamma_i H_C} \ \dots \  \underbrace{e^{-i\beta_1 H_B}e^{-i\gamma_1 H_C}}_{first \ layer}\ket{+}^{\otimes n}
\end{equation}
where the cost layer and mixer layer are put alternatively, variational parameters $\gamma_i$ and $\beta_i$ may differ for each component.

% Interestingly, a trick is used, thinking of a quantum circuit in terms of a Hamiltonian. This indicates that a gate can be thought of as a physical implementation obtained by carrying out time evolution under a meticulously designed Hamiltonian. Here the unitary time evolution operator is defined as:
% \begin{equation}
%     U(H, t) = e^{-iHt/\hbar}
% \label{equ: QAOAtimeEvolution}    
% \end{equation}
% where $H$ is a Hamiltonian and scalar $t$ is time.

% To implement an approximate time-evolution unitary by a quantum circuit, Equation~\ref{equ: QAOAtimeEvolution} could be reformulated as:
% \begin{equation}
%     U(H, t, p) = \prod\limits_{j = 1}^p\prod\limits_{k = 1}^{n}e^{-iH_kt/p}, \quad H = \sum\limits_{k = 1}^nH_k,
% \end{equation}
% where $U$ approaches $e^{-iHt}$ as $p$ becomes larger.

% \begin{equation}
%     \ket{\psi(\gamma, \beta)} = e^{-i\beta_n H_B}e^{-i\gamma_n H_C}\ \dots \ e^{-i\beta_i H_B}e^{-i\gamma_i H_C} \ \dots \ e^{-i\beta_1 H_B}e^{-i\gamma_1 H_C}\ket{\psi(0)}
% \end{equation}

% \begin{equation}
%     U(\gamma, \beta) = e^{-i\beta_p H_B}e^{-i\gamma_p H_C}\ \dots \ e^{-i\beta_i H_B}e^{-i\gamma_i H_C} \ \dots \ e^{-i\beta_1 H_B}e^{-i\gamma_1 H_C}
% \end{equation}

% Let $F_p(\gamma, \beta)$ be the expectation value of $H_C$ in this state:
% \begin{equation}
%     F_p(\gamma, \beta) = \bra{\gamma, \beta}H_C\ket{\gamma, \beta}.
% \end{equation}

QAOA runs the circuit with initial parameters $(\gamma, \beta)$ on a quantum computer for a given problem Hamiltonian and measures in the computational basis to determine the expectation value $H_C$:
\begin{equation}
    F_p(\gamma, \beta) = \bra{\gamma, \beta}H_C\ket{\gamma, \beta}.
\end{equation}
After that, QAOP employs classical techniques to optimize the parameters $(\gamma, \beta)$ for minimizing or maximizing the $F_p$. Once the circuit (i.e., parameters $(\gamma, \beta)$ ) is optimized, the state produced by quantum computers is measured again. After this procedure is repeated a sufficient number of times, measuring the final state produced on quantum computers yields approximate solutions to the optimization problem. QAOA is a layerized quantum circuit, which is visualized in Figure~\ref{fig: QAOA}. QAOA is a highly-regarded method for solving combinatorial optimization problems on Noisy Intermediate-Scale Quantum (NISQ) devices. This is because it employs shallow quantum circuits, coupled with the variational nature that provides robustness to systematic errors, making it a leading candidate to achieve quantum advantage.

The general structure of QAOA is a \textit{Variational Quantum Algorithm} (VQA) \cite{Cerezo_2021} that provides a comprehensive framework capable of addressing an array of problems. Another famous application of such a framework is the Variational Quantum Eigensolver (VQE) \cite{peruzzo2014variational}. VQA is a hybrid quantum-classical optimization algorithm, which could be considered as the quantum counterpart of machine learning methods such as neural networks. This algorithm uses a quantum computer to estimate an objective function having parameters while the parameters of this function are updated by leveraging the power of classical optimizers. This methodology offers the benefit of maintaining a shallow quantum circuit depth, thereby mitigating noise, a stark contrast to quantum algorithms crafted for the fault-tolerant era. Emerging as a vital strategy for achieving quantum advantage on NISQ devices, VQAs get a lot of attention due to their favorable scalability with increasing qubits and their ability to operate without high fault tolerance. Despite their potential in obtaining near-term quantum advantage, VQAs still grapple with significant challenges including trainability, accuracy, and efficiency.

In the following sections, this survey undertakes a thorough examination of the current body of quantum-related literature within the realm of databases. Through this exhaustive exploration, our aim is to stimulate enthusiasm within the database community, thereby encouraging its members to foster the growth of a new sub-community focused on quantum computing in databases and actively contribute to its advancement.

\begin{table}[tbp]
\centering
\setlength{\extrarowheight}{0pt}
\addtolength{\extrarowheight}{\aboverulesep}
\addtolength{\extrarowheight}{\belowrulesep}
\setlength{\aboverulesep}{0pt}
\setlength{\belowrulesep}{0pt}
\caption{Comparison among Quantum Database Search}
\label{tab:quantumDatabaseSearch}
\begin{tabular}{m{7em}<{\centering}m{17em}<{\centering}m{10em}<{\centering}}
\midrule
% \rowcolor{lightgray!20}
\rowcolor[gray]{.9}
\textbf{Time} & \textbf{Works} & \textbf{Complexity of query}
\\ \toprule
1996 / 1997 &\textit{Grover's algorithm}\cite{DBLP:conf/stoc/Grover96,Grover_1997}&  $O(\sqrt{N})$  \\ 
\hline

1997 & \textit{Grover} \cite{Grover_1997Single_Query} &  1  \\ 
\hline

1998 & \textit{Terhal and Smolin} \cite{Terhal_1998} &  1  \\ 
\hline

1998 & \textit{Boyer et al.} \cite{Boyer_1998} &  $O(\sqrt{N/t})$ \\ 
\hline

2001 & \textit{Patel} \cite{patel2001quantum} &  $O(log_4 N)$  \\ 
\hline

2004 & \textit{Imre and Bal{\'{a}}zs} \cite{DBLP:journals/computing/ImreB04} &  $O(\sqrt{N})$ or 1  \\ 
\hline

2002 / 2007 & \textit{Tsai et al.} \cite{1032254} / \textit{Ju et al.} \cite{Ju2007QuantumCD} &  $O(\sqrt{N})$  \\ 
\hline

\bottomrule
\end{tabular}
\end{table}

\section{Quantum computing-enabled technology}\label{sec:quantum-based}

In this section, we have explored the use of quantum computing-enable techniques to optimize databases in different aspects of problems. These techniques leverage the unique capabilities of quantum computing to enhance existing database technologies and develop new ones. Furthermore, there is strong anticipation for a rapid increase in the capacity of quantum computers in the near future, accompanied by the availability of on-site quantum computers. This significant advancement in quantum computing technology will create enhanced prospects for the application of the techniques elucidated in this chapter.

\subsection{Quantum Database Search}

% ------------------------------------------------------------------------------------------

% This sections reviews Grove search algorithms and its extension for database search.

% Gongsheng write this part. References include at least

%  \cite{DBLP:conf/stoc/Grover96,DBLP:journals/computing/ImreB04,patel2001quantum}

% ----------------------------------------------------------------------

Shor's algorithm, a quantum algorithm, for finding the prime factors of an integer in polynomial time shows that quantum computers do have an advantage in addressing some problems for which there was no known efficient classical algorithm. The proposal of Grover's algorithm \cite{DBLP:conf/stoc/Grover96, Grover_1997} further aroused a lot of excitement from computer scientists, especially from researchers in databases. This is because, with Grover's algorithm, there is a high possibility that people could know whether or not unsorted databases containing $N$ records have one satisfying particular property by $O(\sqrt{N})$ operations instead of checking records one by one until finding the desired one. Based on this benefit of Grover's algorithm, people have made the following efforts in the hope of improving database query performance. In this subsection, we will introduce current research in this area and summarize those works in Table~\ref{tab:quantumDatabaseSearch}. Besides, there are some works about quantum searching, and interesting readers may refer to  \cite{PhysRevA.60.2746, Grover1997QuantumCC, Long1999PhaseMI, arxiv.quant-ph/9911004, PhysRevA.64.022307, arxiv.quant-ph/0107013, PhysRevA.65.052322, Biron1998GeneralizedGS, PhysRevA.60.2742, PhysRevA.63.012310, 996861, Grover2002TradeoffsIT, AZUMA_2000, Braunstein2000SpeedupAE, Guilu1999ArbitraryPR, Long1999PhaseMI, LONG2002143}.

\textit{\textbf{Grover}} \cite{Grover_1997Single_Query} puts forth a quantum algorithm locating a marked item with only a single query in a database having $N$ items, where a query is defined as any question to the database to which the database has to return a (YES/NO) answer. The algorithm is implemented using a collection of $m$ (i.e., $\Omega(NlogN)$) independent subsystems, each with an $N$-dimensional state space like the system of the original Grover's algorithm \cite{DBLP:conf/stoc/Grover96,Grover_1997}. In detail, the algorithm starts with a tensor product of $m$ identical quantum subsystems and initializes all subsystems to the superposition over all $N$ states. Then it boosts the amplitude of the target state(s) in each subsystem. Finally, it chooses the item pointed to by most subsystems as the desired one.

% \textit{\textbf{Grover}} \cite{Grover_1997Single_Query} proposes an algorithm to locate the marked item in a single query (a query is defined as any question to the database to which the database has to return a (YES/NO) answer). This algorithm is implemented by using enough subsystems, where each subsystem has an $N$ dimensional state space like the system of the original Grover's algorithm \cite{DBLP:conf/stoc/Grover96,Grover_1997}, and the number of subsystems (denoted as $m$) needs to be $\Omega(NlogN)$. In detail, the algorithm starts with a tensor product of $m$ identical quantum subsystems and initializes all subsystems to the superposition over all $N$ states. Then it boosts the amplitude of the target state(s) in each subsystem. Finally, it chooses the item pointed to by most subsystems as the desired one.

%disadvatage:(ii)	As mentioned previously, the same algorithm applies when more than one item is marked, with the caveat that the N number of marked items is less than	N/4. There are two reasons for this limitation.

% --------------------------------------------
%----------------------------------------------
%-----------------------------------------------
% --------------------------------------------
%----------------------------------------------
%-----------------------------------------------

%Single quantum querying of a database  1998

\textbf{\textit{Terhal and Smolin}} \cite{Terhal_1998} introduce a family of efficient quantum algorithms, drawing inspiration from Bernstein and Vazirani's parity problem, that is capable of retrieving the entire contents of a quantum database with only a single query. The class of algorithms encompasses binary search problems and coin-weighing problems. In problems, a database $Y$ (having $N$ items) is represented as an $n$-bit string $y$ with Hamming weight one, of which a single item is marked. The answer to queries (denoted as $n$-bit strings $x$) returned by the database is : $a(x, y) = x \cdot y \equiv (\sum_{i = 1}^n x_i y_i) mod 2$,
where $x_i$ and $y_i$ are the $i^{th}$ bits of $x$ and $y$. The quantum database interacts with two input registers: register $X$, which houses the query state denoted as $\ket{x}$, and register $B$, which serves as an output register of dimension $A$ and initially comprises the state $\ket{b}$. The process of querying the database can be defined by the operation $R_y: \ket{x,b} \rightarrow \ket{x, [b + a(x, y)] mod A}$, where $R_y$ signifies a classical reversible transformation. This transformation maps basis states to other basis states (thus, it can be considered a permutation matrix), and its operation is contingent on the contents of the respective database. Within this operation, $a(x, y)$ corresponds to the answer to a given query $x$, considering the current state $y$ of the database. For the coin-weighting problem, it aims to identify defective coins. A set of $n$ coins is regarded as a database and represented as $y$, where $y_i = 1$ means that coin $i$ is defective. One weighing is one query $x$, where $x_i = 1$ specifies whether coin $i$ is included in the set to be weighed. With the Bernstein-Vazirani algorithm, authors first create a query state $\ket{\psi}$. After a query, a new state $\ket{\psi_y}$ is obtained, which contains the information of $a(x,y)$. Then, a Hadamard transform is performed on the query register to get the final state, thus retrieving $y$ within a single query. A similar process is applied to solving the binary search problem with classical encoding schemes (Huffman coding).

\textit{\textbf{Boyer et al.}} \cite{Boyer_1998} proposes an algorithm suitable for situations where the number of items (the desired items), denoted as $t$, in an $N$-items table is not known ahead of time. First, one interesting thing is given in this paper, an improvement in complexity is expected if the algorithm stops after $j$ iterations, observes the register, and starts all over again in case of failing to find target items. Then, based on this, the main idea of the proposed algorithm is to repeatedly execute Grover's algorithm with a predetermined number of iterations $j$ and constantly increase $j$ to achieve the goal of finding the target and, more importantly, to reduce the total number of iterations. The specific implementation of the algorithm is as follows, which could solve the problem with a $O(\sqrt{N/t})$ complexity when $1 \leq t \leq 3N/4$. The scenario when $t$ exceeds $3N/4$ can be efficiently resolved through classical sampling, whose complexity is in the constant expected time. On the other hand, the situation when $t$ equals 0 is handled by incorporating an optimal time-out. This integration allows the assertion that no solutions exist in a time complexity of $O(\sqrt{N})$. Finally, inspired by Shor's quantum factorization algorithm, the authors combine its technologies with Grover's algorithm to sketch one basic idea for approximately counting the number of solutions (i.e., $t$). More details are given in the paper \cite{10.1007/BFb0055105, Brassard_2002}.

\begin{enumerate}\small
    \item Initialize $m = 1$ and set $\lambda$ = 6/5 (pick $\lambda$ randomly and 1 $\leq$ $\lambda$ $\leq$ 4/3);
    
    \item Select an integer $j$ uniformly at random and $0 < j < m$;
    
    \item Apply $j$ iterations of Grover’s algorithm starting from the initial superposition state;
    
    \item Get the outcome $i$ from the register ($i$ indicates the location of the item in the table);
    
    \item If $T [i] = x$ (the desired item), exit.
    
    \item Otherwise, set $m = min(\lambda m, \sqrt{N})$ and go back to step 2.

\end{enumerate}

\textit{\textbf{Patel}} \cite{patel2001quantum} introduces a factorized quantum search algorithm as a means to efficiently locate a desired object within an unordered database. This algorithm could reach a querying complexity of $O(\log_4 N)$, resembling the DNA replication assembly process. The idea behind this is sorting could facilitate searches while the acceleration realized through sorting can be understood as factorization of the search process, which is especially suited to quantum databases having superposition. This is because the order is redundant when states are superposed in quantum databases. To use factorization to enhance the search performance, the algorithm initiates by digitizing the contents of the database. This process equips with an alphabet having $a$ letters. Then it uses a string having $n$ letters to label $N (N = a^n)$ items in databases. Here, $a$ could be set to 4. Next, it starts with the uniform superposition of all $N$ states. With digitization, the algorithm is able to check one letter of integer labels at a time. The method is through the collection of factorized oracles $F_i(x_i)$ not using a global oracle $F(x)$, where $F_i(x_i)$ takes the $i^{th}$ letter of the label as input and output $-1$ (i.e., the letter matches the corresponding letter of the wanted string) or $+1$ (i.e., does not match). Besides,  the reflection operation $R_i$ follows $F_i$ to reflect all the amplitudes about their average. After $R_iF_i$, the algorithm doubles the amplitudes of all those states satisfying $F_i(x_i) = -1$ and decreases all the other amplitudes to zero. After that, this algorithm utilizes the projection/measurement operator $P_i$ ($P_i = I_1 \otimes \dots (I_i)_{fix} \otimes \dots \otimes I_n$) to remove from the Hilbert space all the states whose $i^{th}$ letter is not what we want. The algorithm sequentially scrutinizes each letter until all the letters are checked and identify the desired item.

\textit{\textbf{Imre and Bal{\'{a}}zs}} \cite{DBLP:journals/computing/ImreB04} propose a generalized Grover operator to enable it to be a building block of a larger quantum system and achieve high accuracy in finding desired results meanwhile letting the algorithm iterate as few times as possible. Specifically, since the state of the input index register of the search algorithm, in practice, may be offered by an arbitrary output state of a former circuit, the initial state of the index register is defined as $\ket{\gamma} \triangleq \sum_{x = 0}^{2^n -1} \gamma_x \ket{x}$ ($\sum_{x = 0}^{2^n -1} |\gamma_x|^2 = 1$) to support arbitrary initial distribution.
Then, the Hadamard transformation ($\mathcal{H}$) is replaced by an arbitrary unitary transformation ($\mathcal{U}$). In the Oracle ($\mathcal{O}$), the probability amplitude of the marked terms in the index register will be rotated with an angle of $\phi$, instead of $\pi$, where $\phi \in [-\pi, \pi]$. Thus, it is denoted with $\mathcal{I_\beta}$ not $\mathcal{O}$.
Next, the controlled phase gate ($\mathcal{P}$) is founded on an arbitrary basis state $\eta$ not on the state $\ket{0}$, denoted with $\mathcal{I_\eta}$. Finally, basis vectors $\ket{\alpha}$ and $\ket{\beta}$ are also redefined. Based on those new definitions mentioned above, the generalized Grover operator ($\mathcal{Q}$) is described by $-\mathcal{U} \mathcal{I}_\eta \mathcal{U}^\dag \mathcal{I}_\beta$.
As for the required number of iterations $l_{MC}$ of the algorithm, it could be derived from the formula (i.e., $\tfrac{\frac{\pi}{2} - |arcsin(sin(\frac{\phi}{2} - \Lambda + \Lambda_\gamma)sin(\Omega_\gamma)|}{\Delta}$) according to the Matching Condition $|\Delta| \leq \tfrac{\pi}{2}$, which could lead the algorithm to sure success measurement. For variables that appear in the $l_{MC}$, different settings will result in different iterations. This paper discusses several cases and shows a setting for $l_{MC} = 1$.
Furthermore, the authors provide a mathematical formula to express the final state obtained after executing $l_{MC}$ iterations, which could be helpful when the algorithm's output is applied as an input to another circuit.

\textit{\textbf{Ju et al.}} \cite{Ju2007QuantumCD} use the quantum Boolean circuits to implement the oracle and the inversion-about average function in Grover's algorithm. The implementation of Grover's algorithm needs to consider two important operations: selective-inversion $I_{\ket{x_0}}$, and inversion-about-average $I_{\ket{\psi_0^{\perp}}}$, where $I_{\ket{x_0}}$ inverts the target record $x_0$ while leaving others unchanged, $I_{\ket{\psi_0^{\perp}}}$ works on the superposition state ($\ket{\psi_0} = \tfrac{1}{\sqrt{2^n}} \sum_{i = 0}^{2^n - 1} \ket{i}$) to invert the components in dimensions that are perpendicular to $\ket{\psi_0}$ while keeping the component in the $\psi_0$ direction unchanged. $I_{\ket{x_0}}$ followed by $I_{\ket{\psi_0^{\perp}}}$ could amplify the probability amplitude of the target. The construction of these two parts utilizes the property of eigenvalue kickback. Besides, from the circuit design perspective, the authors present applications such as a phone-book-like database searching problem and breaking a symmetric cryptosystem problem, which is suitable to be solved using Grover's algorithm. Finally, the authors summarize these problems and offer a template of quantum circuits for solving the following situation: Give a one-way function $f: \{0, 1\}^n \longrightarrow \{0, 1\}^m$ and a $m$-bit integer $M$, the aim is to find an $n$-bit integer $x$ for satisfying $f(x) = M$. This paper is an extension of \cite{1032254}.

\paragraph{\textbf{Discussion}}

By making good use of the superposition of states and multiple operations at the same time, Gover proposes a quantum search algorithm \cite{DBLP:conf/stoc/Grover96, Grover_1997} to demonstrate that we could identify the desired element, in an unordered database containing $N$ items, within $\sqrt{N}$ steps. This attracts many people's attention and interest. The reason is there is a common and long-term perception that we need to take $O(N)$ steps to check $N$ items for answering this question. With $\sqrt{N}$ steps, it is counter-intuitive. But this is an important finding. From an engineering point of view, many problems can be formulated as a database searching process. If we solve this problem with as few queries as possible, the performance of many applications could be improved tremendously, like database systems. Considering the potential of quantum computing, Grover further proposes an algorithm \cite{Grover_1997Single_Query} to locate the marked item in a single query. But it should be noted that the number of marked items needs to be less than $N/4$ when the algorithm is applied to queries pertaining to multiple marked items, as once there are too many marked items, the probability between marked items and other items cannot be distinguished. Besides, while the algorithm only requires a single query, the preprocessing and postprocessing procedures are still $\Omega(NlogN)$. After that, many works have been presented to improve Grover's algorithm (like \cite{Boyer_1998,patel2001quantum, DBLP:journals/computing/ImreB04}) or redesign algorithms for finding the marked items with one query (like \cite{Terhal_1998}). \textit{Tsai et al.} \cite{1032254} and \textit{Ju et al.} \cite{Ju2007QuantumCD} provide a quantum circuit template to search the solutions of the problems formulated as a certain class of one-way functions. However, the authors regard that compared to the high efficiency of solving one-way functions with clear mathematical function correspondence, this template (or Grover's algorithm) is impractical for solving one-way functions without it. For example, in the phone-book-like database searching problem, each record $(name, number)$ about $Phonebook(name) = number$ is different and cannot be formulated effectively. But, in a cryptosystem, $E(k, P) = C$ ($P$ and $C$ mean plaintext and ciphertext respectively, $k$ is key.), the key-value pair $(P, C)$ has a very specific formulation. In addition, the problem, like the phone-book-like database searching problem, has to spend some time $O(N)$ to build the database before the search process. And the database needs to be fully reconstructed after each query due to the collapse of quantum states. The most important is the classical database could locate the desired record in $O(logN)$ with indexes. Furthermore, we think how to use quantum computing in the field of databases is still not so clear, which remains an attractive goal for future research. Maybe we could begin by integrating these works with efficient database search algorithms that employ the database's specific properties to improve the database's performance.

\begin{table}[tbp]
\centering
\setlength{\extrarowheight}{0pt}
\addtolength{\extrarowheight}{\aboverulesep}
\addtolength{\extrarowheight}{\belowrulesep}
\setlength{\aboverulesep}{0pt}
\setlength{\belowrulesep}{0pt}
\caption{Comparison among Database Manipulation}
\label{tab:quantumDatabaseManipulation}
\begin{tabular}{m{4em}<{\centering}m{13em}<{\centering}m{18em}<{\centering}}
\midrule
% \rowcolor{lightgray!20}
\rowcolor[gray]{.9}
\textbf{Time} & \textbf{Works} & \textbf{Operations}
\\ \toprule
1997 & \textit{Cockshott} \cite{9712025} &  Selection, Projection, Join  \\ 
\hline

2007 & \textit{Younes} \cite{arxiv.0705.4303} &  Insert, Update, Conditional Operations, Backup, Restore  \\ 
\hline

2007 & \textit{Liu and Long} \cite{arxiv.0710.3301} &  Delete  \\ 
\hline

2010 & \textit{Gueddana et al.} \cite{Gueddana2010OptimizedMF} &  Insert, Delete, Select  \\ 
\hline
\bottomrule

2008 & \textit{Pang et al.} \cite{Pang2008QuantumSA} &  Intersection  \\ 
\hline

2012 & \textit{Salman and Baram} \cite{Salman2012QuantumSI} &  Intersection  \\ 
\hline

2020 & \textit{J{\'{o}}czik and Kiss} \cite{DBLP:conf/adbis/JoczikK20} & Intersection, Difference, Union, Projection  \\
\hline

\bottomrule
\end{tabular}
\end{table}

\subsection{Database Manipulation}

It is widely recognized that database management systems (DBMSs) possess the capability to manipulate data. The operations of manipulating data include but are not limited to \textit{insert}, \textit{delete}, \textit{select}, \textit{project}, etc. In this subsection, we will introduce current research on applying quantum computing in this area and summarize those works in Table~\ref{tab:quantumDatabaseManipulation}.

%interact with end users, applications, and the database itself to capture and analyze the data. 

%Quantum Relational Databases  1997

\textbf{\textit{Cockshott}} \cite{9712025} shows how the basic operations (selection, projection, and join) of relational databases are implemented on a quantum computer. \textit{SELECT}: the primary key selection could be executed with means of Grover's algorithm \cite{DBLP:conf/stoc/Grover96, Grover_1997}.
\textit{PROJECT}: relational database projection will be implemented by only keeping the qubits coding for the domains onto which the relation is projected while discarding the rest qubits. 
\textit{JOIN}: the join operation can be implemented by combining a similarity operator $\approx$, a combining operator $\oplus$, and Grover’s algorithm \cite{DBLP:conf/stoc/Grover96, Grover_1997} (i.e., $r(\oplus \bowtie \approx)s$, where $\bowtie$ is the join functional, $\oplus$ is dyadic combination operator, $\approx$ is similarity operator, $r$/$s$ is a relation.).

% ----------------------------------------------------------
% ----------------------------------------------------------
% ----------------------------------------------------------
% ----------------------------------------------------------

% --------------------------------------------
%----------------------------------------------
%-----------------------------------------------
% --------------------------------------------
%----------------------------------------------
%-----------------------------------------------

%[6]	A. Younes: Database manipulation on quantum computers, ArXiv Quantum Physics e-Prints, Cornell Univ., Ithaca, NY, May. 2007

\textbf{\textit{Younes}} \cite{arxiv.0705.4303} propose some definitions of basic operations about a Quantum Query Language (QQL) to manipulate a database file. \textit{INSERT}: The utilization of controlled Hadamard gates to implement inserting records in sequence into a superposition is presented, which includes inserting one at a time and inserting several at a time. \textit{UPDATE}: The utilization of CNOT gates \cite{younes2003automated} to implement the permutation operator for updating records is presented, which could help insert operations escaping the previous rules of inserting in sequence.
\textit{Conditional Operations}: Select certain sets of records by Boolean functions $f$. Then a conditional operation $U$ (an arbitrary operation) could be performed on the intersection of the selected records according to some global conditions.
\textit{BACKUP} and \textit{RESTORE a BACKUP} are proposed based on the oracle and partial diffusion operators.

\textbf{\textit{Liu and Long}} \cite{arxiv.0710.3301}  offer a quantum algorithm that enables the deletion of a marked item from an unsorted database through a single query without any knowledge of the marked state. This quantum deletion algorithm is implemented by calling iteratively the following subroutine: \ding{182} Execute a conditional phase shift $e^{i\phi}$ on every basis state except the marked state $\ket{\tau}$; \ding{183} Execute the Hadamard transformation on $n$ qubits; \ding{184} Execute a conditional phase shift $e^{i\phi}$ on the $\ket{0}$ state while keeping other basis states unchanged; \ding{185} Execute the Hadamard transformation on $n$ qubits again. They show that with a fixed phase angle $\phi$ ($\pi/3$), the marked state is near completely deleted after executing the proposed deletion algorithm. Furthermore, this deletion operation is periodic and has a period of 3.

\textbf{\textit{Pang et al.}} \cite{Pang2008QuantumSA} propose a quantum algorithm that employs a combination of Grover's algorithm, classical memory, and classical iterative computation to obtain the intersection of two sets. This combination could achieve a reduction in algorithmic design complexity. The algorithm comprises two parts: a subroutine finding an element in set $X = A \bigcap B$, akin to the BBHT algorithm proposed in \cite{Boyer_1998}, where the main difference is it employs general Grover iteration $G_{general}$ here; and a second part that obtains $X = A \bigcap B$ by repeatedly invoking the subroutine until no output is yielded from the subroutine, where this process involves database updates to prevent finding the same element when repeatedly calling the subroutine.

%12' The Building and Optimization of Quantum Database 抄袭
%yuxing: should belong to query optimization or the implementation of a real quantum database, e.g., how DML(update read) is encoded in quantum computing
% Gongsheng: Okay! Thank you! I will check it.

%Optimized methods for inserting and deleting records and data retrieving in quantum database 2010

\textbf{\textit{Gueddana et al.}} \cite{Gueddana2010OptimizedMF} propose a quantum database design for a relational database consisting of multiple tables, each of which could be represented by a circuit, such as a CNOT-based circuit \cite{younes2003automated}. The quantum database associated with the relational database is considered to be a collection of quantum circuits. The authors then offer circuits-based implementation of querying quantum tables, inserting records, and deleting records from a quantum table, specifically focusing on retrieving records from naturally joined quantum tables.

\textbf{\textit{Salman and Baram}} \cite{Salman2012QuantumSI} present an algorithm for selecting a member of the intersection set $X = A \bigcap B$, which is predicated on the utilization of the oracles $f_A$ (where $f_A(x) = 1$ if $x \in A$, otherwise $f_A(x) = 0$), $f_B$ (where $f_B(x) = 1$ if $x \in B$), the intersection oracle $f_{A \bigcap B}(x)$, and a Toffoli gate. Subsequently, the algorithm employs Grover's algorithm \cite{DBLP:conf/stoc/Grover96, Grover_1997} to retrieve an element from the intersection set $X$.

\textbf{\textit{J{\'{o}}czik and Kiss}} \cite{DBLP:conf/adbis/JoczikK20} propose a set of four quantum algorithms that utilize Grover's algorithm \cite{DBLP:conf/stoc/Grover96, Grover_1997} to facilitate fundamental set operations, including intersection, difference, union, and projection. The first algorithm involves defining an appropriate Oracle function $O_f$ to identify the intersection set $X$ of sets $A$ and $B$ (i.e., $O_f(X) = 1$ when $X = A \bigcap B$). Because $A \bigcap B$ belongs to the power set of $A$ (i.e., $A \bigcap B \in \mathcal{P}(A)$, $|\mathcal{P}(A)| = 2^n$), Grover's algorithm can be used to search for $A \bigcap B$ on the quantum bits, where each bit represents an element of $A$. The second algorithm gets the set difference (e.g., $A \setminus B$) by leveraging the previous intersection algorithm to obtain the result (i.e., $A \setminus B = A \setminus (A \bigcap B)$, the complement set of $(A \bigcap B)$ relative to $A$). The third algorithm obtains the union $A \bigcup B = (A \setminus B) \bigcup B$. It is evident that $A \setminus B$ and $B$ are two disjoint sets, so $A \bigcup B$ can be obtained by simply appending $A \setminus B$ to $B$. The final algorithm retrieves the projection and comprises two parts: the removal of duplicate attributes, accomplished by converting a multiset to a set, and the acquisition of specific column elements, implemented through Grover's algorithm.

\paragraph{\textbf{Discussion}}

The quantum computer could offer the ability to conduct data processing simultaneously. 
To obtain the benefit from the potential acceleration of quantum computers over classical computers is a crucial purpose for manipulating a database system on a quantum computer \cite{arxiv.0705.4303}. This is because the database field tends to require an extensive amount of storage and process time. Although some tentative work has been proposed so far\cite{9712025, arxiv.0705.4303, arxiv.0710.3301, Pang2008QuantumSA, SHIMAN20121602, DBLP:conf/adbis/JoczikK20}, it is evident that the primary key selection performance of \cite{9712025} is inferior to the classical database with an index on the primary key. For the work \cite{arxiv.0705.4303}, it presents some operations like insert, update, backup, etc., but the delete, just given in a suggested way, still needs further thinking. As for the work \cite{arxiv.0710.3301}, it proposes how to delete a marked item from a database. However, circuit implementation is also neglected like the previous two works. \cite{Gueddana2010OptimizedMF} shows how a quantum database associated with a relational database could be handled through the quantum circuit. Although it gives a circuit implementation of generic SQL-like queries, there are still various database manipulating operations waiting to figure out. 

\cite{Pang2008QuantumSA, Salman2012QuantumSI} propose a quantum algorithm for the intersection operation, which is significant because it is the ground of many fields (e.g., databases). But it does not show how to get other set operations (e.g., union). \cite{DBLP:conf/adbis/JoczikK20} provides the algorithms for implementing several set operations and employs IBM-Q Experience platforms to evaluate them. But since all set operations are inherited from the set intersection implementation with Grover’s algorithm, it must go through substantial possible elements (the power set of the smaller input set) for checking. Therefore, just employing the presented algorithms to implement one single operation is not recommended.

% Inserting, updating and deleting values from superposed states were proposed by A. Younes et al. [6]; however, circuit implementation and evolution of the probability amplitude were neglected in his work. 
% （comments from Paper "The Building and Optimization of Quantum Database"）

%A way to delete certain records from the database simultaneously has been suggested which still need special attention as a separate problem.

% Manipulating a database system on a quantum computer is an essential aim to benefit from the promising speed-up of quantum computers over classical computers in areas that take a vast amount of storage and processing time such as in databases. 

% This gives the ability to perform the data processing, that usually takes a long processing time on a classical database system, in a simultaneous way on a quantum computer. 

% In this paper, the basic operations for manipulating the data in a quantum database will be defined, e.g. INSERT, UPDATE, DELETE, SELECT, backing up and restoring a database file. 

% ----------------------------------------------------------
% ----------------------------------------------------------
% ----------------------------------------------------------
% ----------------------------------------------------------

\subsection{Database Query Optimization}

% This section discusses how to use quantum computer to speed up the query processing for databases.

Query optimization involves identifying an optimal way to execute a given query, which typically entails exploring extensive search spaces to resolve it. This problem has attracted considerable attention over several decades and has been extensively investigated \cite{10.582099}. Nonetheless, query optimization becomes more complex as database systems grow increasingly complicated, making exploring unconventional approaches attractive. In this subsection, we shall present contemporary studies on integrating quantum computing into query optimization and provide a summary of these endeavors in Table~\ref{tab:quantumDatabaseQueryOptimization}.

\begin{table}[tbp]
\centering
\setlength{\extrarowheight}{0pt}
\addtolength{\extrarowheight}{\aboverulesep}
\addtolength{\extrarowheight}{\belowrulesep}
\setlength{\aboverulesep}{0pt}
\setlength{\belowrulesep}{0pt}
\caption{Comparison among Database Query Optimization}
\label{tab:quantumDatabaseQueryOptimization}
\begin{tabular}{m{3em}<{\centering}m{11em}<{\centering}m{11em}<{\centering}m{10em}<{\centering}}
\midrule
% \rowcolor{lightgray!20}
\rowcolor[gray]{.9}
\textbf{Time} & \textbf{Works} & \textbf{Quantum System} & \textbf{Problem}
\\ \toprule
2016 & \textit{Trummer and Koch} \cite{DBLP:journals/pvldb/Trummer016}  &  Annealer & MQO \\ 
\hline

2021 & \textit{Fankhauser et al.} \cite{arxiv.2107.10508}  &  Gate-Based  & MQO \\ 
\hline

2022 & \textit{Sch{\"{o}}nberger} \cite{DBLP:conf/sigmod/Schonberger22}  &  Gate-Based \& Annealer & MQO \& Joining Order\\ 
\hline

2023 & \textit{Sch{\"{o}}nberger et al.} \cite{sigmod2023}  &  Gate-Based \& Annealer & Joining Order\\ 
\hline

2023 & \textit{Nayak et al.} \cite{10.1145/3579142.3594298} & Gate-Based \& Annealer & Joining Order\\ 
\hline

2023 & \textit{Winker et al.} \cite{winker2023quantum} & Gate-Based & Joining Order\\ 
\hline

2023 & \textit{Sch{\"{o}}nberger et al.} \cite{DBLP:conf/vldb/SchonbergerTM23} & Gate-Based \& Annealer & Joining Order \\ \hline

2023 & \textit{Gruenwald et al.} \cite{DBLP:conf/vldb/GruenwaldWCGG23} & Gate-Based & Index Tuning\\ 
\hline

\bottomrule
\end{tabular}
\end{table}

% --------------------------------------------
%----------------------------------------------
%-----------------------------------------------
% --------------------------------------------
%----------------------------------------------
%-----------------------------------------------

% \textit{Trummer and Koch} \cite{DBLP:journals/pvldb/Trummer016}

%Multiple Query Optimization on the D-Wave 2X Adiabatic Quantum Computer 2016

\textbf{\textit{Trummer and Koch}} \cite{DBLP:journals/pvldb/Trummer016} propose an algorithm that employs an adiabatic quantum computer to address the Multiple Query Optimization (MQO) problem. MOQ involves selecting a query plan combination that minimizes the cost of executing a corresponding set of queries. To accomplish this task, the algorithm starts with converting an MQO problem instance into a Quadratic Unconstrained Binary Optimization (QUBO) instance. This is because the quantum annealer can only handle QUBO problems. Due to a QUBO problem only allowing binary variables, the algorithm offers binary variables $X_p$ for each query plan $p$ to show whether $p$ is performed. Since the optimal solution to a QUBO problem is to minimize a quadratic formula (aka, energy formula), the algorithm converts the provided MQO instance into an energy formula on those binary variables. Now, the MQO problem is transformed into the variable assignment problem. So far, the above process is called logical mapping, and the energy formula is the logical energy formula. In this logical mapping, the variables $X_p$ are called the logical variables, which means that variables $X_p$ can not yet be expressed by the quantum annealer's qubits. Next, the algorithm picks a bunch of physical qubits to express separately logical variable $X_p$ so that it can put the weights on single qubits and the coupling strengths between qubits, thus reformulating the logical energy formula as a physical energy formula. And this process is named physical mapping. Subsequently, the algorithm utilizes the quantum annealer to minimize the physical energy formula by finding an optimal value assignment, followed by a step-by-step passback of the solution provided by the quantum annealer. And it uses the returned solution to obtain the answer to the logical energy formula. Finally, the algorithm uses the obtained answer to address the original MQO problem.

\textbf{\textit{Fankhauser et al.}} \cite{arxiv.2107.10508} introduce an approach to finding quasi-optimal solutions for the MQO problem on a gate-based quantum computer, which is proposed based on the Quantum Approximate Optimization Algorithm (QAOA). The proposed approach is made up of two components: firstly, the exploration of the search space utilizing parametrized quantum computing, and secondly, the optimization of the parameters through classical computing techniques employing heuristic strategies \cite{zhou2020quantum}. In detail, the algorithm first encodes the MQO problem as a classical cost function $\mathcal{F}_\mathcal{C}$ and then reformulates $\mathcal{F}_\mathcal{C}$ into a quantum cost function $\mathcal{F}_\mathcal{Q}$. Next, the algorithm initializes the parameters $\beta$ and $\gamma$. The third step sets the quantum circuit in superposition and applies parameterized function $\mathcal{F}_\mathcal{Q}(\beta,\gamma)$ in $p$ steps inside the quantum circuit. After that, the algorithm measures the quantum circuit to get a solution to the MOQ problem. Then it employs the classical computer to optimize the $\beta$ and $\gamma$ after each measurement (they will be used in the above third step again) to achieve the goal of producing the least-cost solution. Finally, the algorithm repeats this process until reaching the threshold.

\textbf{\textit{Sch{\"{o}}nberger}} \cite{DBLP:conf/sigmod/Schonberger22} conducts an exploration into the resolution of the MQO problem using QAOA on a 27-qubit gate-based quantum computer, engaging in a thorough analysis of MQO problems with varying numbers of plans (with up to 24) per query. In addition, a multi-stage reformulation of the join ordering problem is proposed, making it run on the gate-based quantum computer and the quantum annealer. In the end, the study concludes by shedding light on the corresponding experimental results. In the end, the study concludes by shedding light on the corresponding experimental results.

\textbf{\textit{Sch{\"{o}}nberger et al.}} \cite{sigmod2023} present a quantum implementation of join ordering optimization based on a reformulation to QUBO problem. In detail, it first encodes the join ordering problem as a Mixed Integer Linear Programming (MILP) problem. Then it transforms the MILP model into a Binary Integer Linear Programming (BILP) formulation. Next, it converts the BILP formulation into a QUBO model appropriate for quantum processing. After that, the QUBO model could be utilized to tackle the join ordering problem by gate-based QAOA and quantum annealing. Based on insights gained from experimental analysis, it is suggested that identified design criteria can be used to craft Quantum Processing Units (QPU) as specialized devices catered to specific applications, such as databases. And the numerical simulations show that through minor physical architectural improvements, the usefulness of a custom co-design QPU for the join ordering can be significantly augmented, which may be a way to accelerate the development of QPUs to achieve near-term utility for databases.

\textbf{\textit{Nayak et al.}} \cite{10.1145/3579142.3594298} propose to model the join order problem as a QUBO problem, where the QUBO formula consists of two parts: the processing cost of the binary join between two relation operands and the penalty for invalid join trees. The goal is to minimize the proposed QUBO formula to retrieve the optimal valid join tree with the smallest costs. This problem is solved with a variety of quantum approaches (e.g., QAOA, VQE, simulated annealing \cite{bertsimas1993simulated}, and quantum annealing) on quantum hardware and simulators.

\textbf{\textit{Winker et al.}} \cite{winker2023quantum} propose a hybrid classical-quantum machine learning algorithm -- predicting efficient join orders from the knowledge learned from past join order -- to solve the join order optimization problem, which runs with a variational quantum circuit (VQC). For this algorithm, authors provide the definitions of encoding (encoding classical data into a quantum state), decoding (decoding quantum result into classical data), reward function, cost, etc. Furthermore, they develop a plugin for PostgreSQL, which permits a trained VQC model to directly integrate into PostgreSQL as a join order optimizer.

%Quantum Optimization of General Join Trees 2023 vldb workshop

% \textbf{new article}
% Therefore, in this paper, we address the drawbacks of the existing JO-QUBO formulation: Rather than transforming an existing encoding into QUBO, thereby inheriting its limitations, we propose a novel QUBO encoding for optimizing general bushy trees, while moreover retaining the ability to identify cross-product solutions. In contrast to many competing JO methods, which either do not consider cross products or restrict the join tree shape, our new encoding enables QPUs to explore the complete, unrestricted search space of the JO problem.

\textbf{\textit{Sch{\"{o}}nberger et al.}} \cite{DBLP:conf/vldb/SchonbergerTM23} address the Join Ordering (JO) problem by natively encoding JO as a QUBO problem rather than transforming existing formulations for JO, which enables QPUs to optimize general, unrestricted join trees while also allowing the use of cross-products, and derive cheap plans. Specifically, to solve JO with optimization methods of QUBO, it is necessary to establish problem formulations that adhere to the prescribed requirements of the QUBO framework. Thus, it needs to identify validity constraints holding for every valid solution, encode these constraints, and set additional terms like costs. Provided that the formulation is appropriately constructed, the minimization of the overall energy formula will yield a specific configuration of variables that corresponds to a valid and optimal solution. For this purpose, the authors first make their encoding for bushy trees generate the join tree itself. Then, the authors translate some relationships like the number of incoming edges, outgoing edges, and no cycles that hold for every possible join tree into penalty terms, to enforce valid assignments for variables representing the edges in the join tree. Due to a lack of a priori information about the relationship between a join pair, authors also use ancillary variables to preserve the depth of all relations and joins in the join tree for preventing the occurrence of cycles in the join tree. Next, in order to assess the quality of a join tree, the authors define variables and penalty terms to encode the logarithmic costs associated with the tree. Finally, the authors approximate the real costs of the join order by leveraging the logarithmic values.

\textbf{\textit{Gruenwald et al.}} \cite{DBLP:conf/vldb/GruenwaldWCGG23} address the challenge of selecting appropriate index configurations for replicated databases, aiming to minimize query processing costs. The problem is complex, especially for large-scale databases, and existing algorithms utilizing heuristic or optimization methods are not efficient enough. While quantum computing has shown promise in various database management areas, no work has explored the divergent design index tuning problem for replicated databases. To fill this gap, the paper proposes a vision of a machine learning-based quantum algorithm. It discusses the issues and presents a classical algorithm that outperforms existing ones, and outlines the transformation to a quantum version.

\paragraph{\textbf{Discussion}}

% Query optimization is finding an optimal way to conduct a provided query by considering the possible query plans, which tends to explore large search spaces for solving this problem. 

% The join order optimization and multi-query optimization, as fundamental query optimization problems, have been extensively investigated for decades \cite{10.582099}. Some works \cite{DBLP:journals/pvldb/Trummer016, arxiv.2107.10508, DBLP:conf/sigmod/Schonberger22, sigmod2023} are currently attempting to find answers to these problems with quantum computing.

In this subsection, we have summarized works \cite{DBLP:journals/pvldb/Trummer016, arxiv.2107.10508, DBLP:conf/sigmod/Schonberger22, sigmod2023, 10.1145/3579142.3594298, winker2023quantum} attempted to find solutions to the query optimization problem with quantum computing. \cite{DBLP:journals/pvldb/Trummer016} tackles the MQO problem by converting the problem into a QUBO problem and shows the experimental results having speedup with three orders of magnitude for the given scenarios. But it only handles relatively small problem instances. And as the number of plans per query increases, it has to reformulate the original problem. \cite{arxiv.2107.10508} uses gate-based quantum systems for tackling query optimization. Although the proposed approach could utilize the quantum device’s qubits more efficiently, its implementation is inferior to methods on a quantum annealing architecture due to the restriction of the number of qubits and fault tolerance of gate-based quantum systems. \cite{DBLP:conf/sigmod/Schonberger22} investigations the method \cite{DBLP:journals/pvldb/Trummer016} with up to 24 plans and studies the scaling behavior of method \cite{arxiv.2107.10508} for transpiled circuits with 27 qubits. Then it proposes a multi-step reformulation for the join ordering problem, but nothing concrete. After that, \cite{sigmod2023} presents a quantum implementation of the join ordering problem. It suggests shaping QPUs as special-purpose devices tailored to database applications, which is more promising based on insights gained from the experimental analysis. However, many problems still exist, like efficient circuit generation against noise and information encoding schemes that might relieve discretization problems \cite{sigmod2023}. And the proposed approach solves QUBO problems by finding the best solution among the left-deep join trees. Against the latter problem, the approach in \cite{10.1145/3579142.3594298} is presented to support more general bushy join trees, which, consequently, enables the determination of the optimal solution among all possible join trees. These days, considering the power of machine learning, \cite{winker2023quantum} proposes a hybrid classical-quantum machine learning algorithm to solve the join order optimization. One of the benefits is that it needs a low number of qubits, which makes optimizing the join order of a practical number of tables possible with current quantum computers. But, the experiments are performed on four tables using simulators. More extensive evaluations are needed. As for \cite{DBLP:conf/vldb/SchonbergerTM23}, it enables Quantum Processing Units (QPUs) to optimize general, unrestricted join trees, albeit at the expense of significantly augmenting the necessary qubit count and substantially expanding the search space. Consequently, this adversely impacts scalability. Different from above,  \cite{DBLP:conf/vldb/GruenwaldWCGG23} discusses the challenge of selecting proper index configurations for replicated databases, minimizing query processing costs.

\subsection{Database Security}

A private search on a database system, which permits the retrieval of an item from the database system without the exposure of the user and the item visited, is one of the most researched topics in database quantum security. 
In traditional database queries, users need to provide explicit search conditions, which may expose their personal information. However, quantum computers can use the quantum random walks algorithm to search the database, which can ensure user privacy and does not require direct disclosure of search conditions. In addition, the quantum entanglement technology used by quantum computers can achieve efficient database queries without exposing the user's search conditions.

Overall, the application of quantum computers in database privacy searches has great prospects, and can provide better protection for personal privacy and enterprise data security.
This section reviews the database security powered by quantum cryptography. 

% Yuxing write this part.

% References include at least \cite{DBLP:journals/compsys/Ozhigov97,DBLP:journals/qip/YangZY15}

% ------------------------------------------------------------

% Gongsheng: Thank you, I see them.
% % belongs to the search problem, sqrt(N) to request N unsorted rows.
% 96' A fast quantum mechanical algorithm for database search
% 97' Quantum computers can search arbitrarily large databases by a single query
% 97' Quantum Relational Databases
% 98' Quantum computers can search rapidly by using almost any transformation
% （A quantum computer has a clear advantage over a classical computer for exhaustive search）
% 98' Single quantum querying of a database

% 97' Protection of information in quantum databases \cite{DBLP:journals/compsys/Ozhigov97}
\textbf{\textit{Ozhigov et al.}} \cite{DBLP:journals/compsys/Ozhigov97} discuss the possible approach to protect information from a database system against a spy with all control system knowledge and having a quantum computer. So to protect the database, a natural quantum computer model is proposed with two components. A traditional part transforms by classical theory (e.g., Turing machine) and a quantum part transforms by quantum mechanics. Then the paper discusses relative diffusion transformations taken used by a quantum database and shows the protection of the database against unauthorized actions and random errors.

% 09' Quantum Information Processing: Algorithms, Technologies and Challenges \cite{reif2009quantum}
\textbf{\textit{Reif et al.}} \cite{reif2009quantum} provide a general survey discussing the algorithms, technologies, and challenges of quantum computing. The authors describe quantum computing and introduce several application fields that involve quantum computing including information process, cryptography, coding theory, and algorithms. To handle such application challenges,  they also introduce several formal quantum computing models, including quantum Turing machines and other Automata, quantum gates, quantum circuits, and computer simulations of quantum computing.  Also, some physical implementations of quantum computing are discussed such as micromolecular quantum computing and nuclear magnetic resonance quantum computing. They provide some quantum cryptography research such as quantum keys and distributed quantum networks. 

% 19' Survey on quantum information security \cite{DBLP:conf/icdis/NjorbuenwuSZ19}
\textbf{\textit{Zhang et al.}} \cite{DBLP:conf/icdis/NjorbuenwuSZ19} recently provide a more specialized survey on quantum information security. With the fast development of quantum computers and quantum algorithms, traditional cryptography algorithms are desired to redesign for security. The author reviews papers theoretically and experimentally in the fields of quantum key distribution, quantum secret sharing, quantum secure direct communication, quantum signature, and private queries. Some research works which are still in the early stage of theoretical studies, such as quantum private comparison and quantum anonymous voting, have been also reviewed.  Some hot research works have been also pointed out for future directions such as the noise problem.

\begin{table}[tbp]
\centering
\setlength{\extrarowheight}{0pt}
\addtolength{\extrarowheight}{\aboverulesep}
\addtolength{\extrarowheight}{\belowrulesep}
\setlength{\aboverulesep}{0pt}
\setlength{\belowrulesep}{0pt}
\caption{Comparison among Database Security}
\label{tab:quantumDatabaseSecurity}
\begin{tabular}{m{3em}<{\centering}m{8em}<{\centering}m{12em}<{\centering}m{13em}<{\centering}}
\midrule
% \rowcolor{lightgray!20}
\rowcolor[gray]{.9}
\textbf{Time} & \textbf{Works} & \textbf{Quantum System} & \textbf{Problem}
\\ \toprule

1997 & \textit{Ozhigov et al.} \cite{DBLP:journals/compsys/Ozhigov97}  & Quantum Turing Machines & Information Protection\\ 
\hline

2009 & \textit{Reif et al.} \cite{reif2009quantum}  & Gate-Based & Information Processing and Cryptography (Survey)\\ 
\hline

2019 & \textit{Zhang et al.} \cite{DBLP:conf/icdis/NjorbuenwuSZ19} & Gate-Based &  Information Security (Survey)\\ 
\hline

\bottomrule
\end{tabular}
\end{table}

\paragraph{\textbf{Discussion} }
One of the most critical filed for quantum computing is \textit{cryptography}, as quantum computing may decode or crack the encrypted key, which may be processed long by traditional computers, within a few seconds or milliseconds. The most research topic in database for cryptography is private search.
The private search on a database system allows a user to retrieve an item from the database system having the retrieved item to be not revealed, known as private information retrieval. Such a query is usually bounded in terms of the database sizes. 

There exist several quantum-based methods 
\cite{DBLP:journals/compsys/Ozhigov97,reif2009quantum,DBLP:conf/icdis/NjorbuenwuSZ19}
% \cite{giovannetti2008quantum,hogg2009private,olejnik2011secure,gao2012flexible,zhang2013private,DBLP:journals/qic/YuQ14,shi2015multi,DBLP:journals/qip/YangZY15,liu2015qkd,DBLP:journals/qip/SunYZ15,yang2016quantum,DBLP:journals/qip/XuSL16,DBLP:journals/qip/YangLCCZS16,wang2016robust,li2016practical,wei2016practical,maitra2017device,yang2017robust,xu2017nearest,wei2017generic,zhou2018quantum,gao2018quantum,yang2019reducing,liu2019qkd,zheng2019practical,hong2019quantum,pei2019practical,gao2019quantum,liu2019quantum,kon2020provably,song2020information,zhao2020development,yang2020multi,wei2020error,yan2020practical,zhou2021quantum,xiao2021quantum,liu2022decoy,wang2022robust} 
proposed for private query problems by guaranteeing the required privacy as well as improving communication complexity and the number of rounds. A more generalized analysis is presented in \cite{giovannetti2010quantum}.
Specifically, some papers 
\cite{liu2019qkd,pei2019practical,wei2020error,liu2022decoy} 
% \cite{gao2012flexible,zhang2013private,liu2015qkd,DBLP:journals/qip/YangLCCZS16,li2016practical,yang2017robust,xu2017nearest,wei2017generic,liu2019qkd,pei2019practical,wei2020error,liu2022decoy} 
focus on the optimization of quantum key distribution (QKD). Some improve to utilizing qubit from a single one \cite{DBLP:journals/qip/YangZY15,yang2017robust} to multiple ones \cite{shi2015multi,DBLP:journals/qip/YangLCCZS16,wang2022robust}. We also provide two surveys related to database security (as shown in Table \ref{tab:quantumDatabaseSecurity}).

% \textbf{yuxing: A more detailed classification will be added to discuss the above many papers}

\subsection{Transaction Management with Quantum Machine}
Transaction management is one of the most important components in database systems. 
The application of quantum computers on transactional databases is gradually developing. Transactional databases refer to database systems with transaction processing capabilities, which can ensure data integrity, consistency, and persistence, thus supporting high-concurrency and high-reliability data processing.
In traditional computer systems, executing transaction processing requires multiple data read-write operations, which consume a lot of time and resources. Quantum computers have parallel processing capabilities and powerful computing power, which can complete a large number of data processing tasks in a shorter time, greatly improving the efficiency and performance of transaction processing.

However, the use of quantum technology and quantum computers is still in the early stages on transaction processing, and there are relatively few studies in this area. The current main work is to model transaction processing as a resource scheduling general problem and then leverage quantum computing techniques to speed it up.
This part introduces some techniques by quantum methods to speed up the transaction process of the concurrent controls.

% % cidr'13 quantum database
% \textbf{\textit{Roy et al.}} \textcolor{red}{duplicate content, see quantum-inspired}\cite{DBLP:conf/cidr/0002KK13} first introduce the concept of a quantum database by providing a solution for OLTP queries in resource allocation problems. The idea is similar to a probabilistic or uncertain database \cite{DBLP:series/synthesis/2011Suciu} but a quantum database keeps uncertainty inside and forces an instantiation. Specifically, it needs to encode SQL queries into resource transactions, a new formalism for the quantum database. Then, an execution model is proposed on top of resource transactions, allowing committed transactions to assign values. The original transactional operations are appropriately executed by maintaining a quantum state, similar to an uncertain state. Finally, a quantum database prototype is deployed as a plugin, providing a common API on top of a MySQL database, and is evaluated to show the overhead as well as the improvement of quantum databases.

% OCJJ'20 Hardware Accelerating the Optimization of Transaction Schedules via Quantum Annealing by Avoiding Blocking
% IDEAS'21 Avoiding blocking by scheduling transactions using quantum annealing
\textbf{\textit{Bittner et al.}} \cite{OJCC_2020v7i1n01_Bittner,Bittner2020} introduce an optimal algorithm by leveraging quantum annealing to optimally schedule transactions over a two-phase locking database such that the performance of a quantum annealer outperforms traditional ones and isolation property is still guaranteed. In general, they model the transaction scheduling problem as QUBO-problems, which consist of binary variables that occur weighted in linear or quadratic terms. Then the problem can be accelerated by quantum annealers via quantum hardware. The evaluation shows that quantum annealing performs constantly and easily outperforms simulated annealing with larger problem sizes.

% IDEAS'21 Optimizing Transaction Schedules on Universal Quantum Computers via Code Generation for Grover's Search Algorithm
\textbf{\textit{Groppe et al.}} \cite{Groppe2021Grover}, similar to \cite{OJCC_2020v7i1n01_Bittner,Bittner2020}, leverage Grover’s search algorithm in solving combinatorial optimization problems to optimally schedule transactions. Firstly, concerning the transaction lengths and conflicts between the transactions, the code generation modular is proposed for the black box function of Grover’s search. And estimating the optimal solutions can further accelerate Grover’s search. The authors also provide a detailed complexity analysis of the comparison between a quantum method and traditional ones in terms of preprocessing and execution time, space, and code length.

% Transaction scheduling \cite{Bittner2020,OJCC_2020v7i1n01_Bittner,Groppe2021Grover} on quantum computers,

\begin{table}[tbp]
\centering
\setlength{\extrarowheight}{0pt}
\addtolength{\extrarowheight}{\aboverulesep}
\addtolength{\extrarowheight}{\belowrulesep}
\setlength{\aboverulesep}{0pt}
\setlength{\belowrulesep}{0pt}
\caption{Comparison among Database Transaction Management}
\label{tab:quantumDatabaseTransactionManagement}
\begin{tabular}{m{3em}<{\centering}m{8em}<{\centering}m{8em}<{\centering}m{13em}<{\centering}}
\midrule
% \rowcolor{lightgray!20}
\rowcolor[gray]{.9}
\textbf{Time} & \textbf{Works} & \textbf{Quantum System} & \textbf{Problem}
\\ \toprule

2020 & \textit{Bittner et al.} \cite{OJCC_2020v7i1n01_Bittner,Bittner2020}  & Annealer & Transaction Schedules to QUBO problems\\ 
\hline

2021 & \textit{Groppe et al.} \cite{Groppe2021Grover}  & Gate-Based & Transaction Schedules speeded up by Grover’s Search\\ 
\hline

\bottomrule
\end{tabular}
\end{table}

\paragraph{\textbf{Discussion}} Transaction processing is one of the core components of database systems. However, there is still little research on quantum computing for transaction processing via our survey. The only research found was to model the transaction processing into a scheduling optimization problem, then used quantum annealer to speed up the optimization (as shown in Table \ref{tab:quantumDatabaseTransactionManagement}). 

Roy et al. \cite{DBLP:conf/cidr/0002KK13} first propose the concept of the quantum database with transactions. They first introduce the concept of resource transactions, which optimally schedules transactions to avoid competition. Followed by such a concept, most later research works leverage different quantum methods, such as quantum annealing \cite{OJCC_2020v7i1n01_Bittner,Bittner2020} and Grover’s search \cite{Groppe2021Grover} to improve the scheduling.

\section{Quantum computing-inspired technology}\label{sec:quantum-inspired}
A motivation for studying quantum computing is the promise that quantum algorithms have the ability to solve some problems better than classical methods. But, as classical algorithm developers learn from innovations in quantum algorithms, several classical algorithms, depending at least partially on developments in quantum computing, have been proposed. These classical algorithms are prepared to run on classical computers, not quantum computers, and are categorized as ``quantum-inspired'' algorithms. Quantum-inspired algorithms are usually inspired by certain principles of quantum mechanics, such as superposition, interference, and coherence, to overcome the classical mechanism's limitation and fulfill a higher performance.
% \textcolor{red}{How about "Quantum Theory Inspired Databases"?}
This part introduces some works based on quantum-inspired methods, which aim to inspire the improvements of ability and performance of databases (see Table~\ref{tab:quantumInspired}).

% But, as classical algorithm developers learn from innovations in quantum algorithms, several classical algorithms have been proposed that depend at least partially on developments in quantum computing.

\subsection{Quantum Inspired Technology for Databases}

\textbf{\textit{Roy et al.}} \cite{DBLP:conf/cidr/0002KK13} propose a middle-tier service to allow resource transactions to commit without assigning specific resource instances (i.e., deferred assignment of values to variables in committed transactions) until observation (i.e., read) by someone for improving the allocation of resources in a dynamic system, whose prototype is implemented as a Java application built over MySQL. 
This middle-tier service could keep track of all possible worlds (i.e., all possible specific resource assignments) for resource transactions. Then, it holds all these possible worlds as its database state. Other transactions can also execute on each possible world and generate another ``forking'' of the state. 
In this way, the database is in a partially uncertain state, named a quantum state. Besides, various transactional operations like reads, writes, the execution of new resource transactions, and grounding (particular value assignment) are offered to update a quantum database state. Finally, it regards this kind of database as a quantum database.

%%%%%%%%%%%%%%%%%%%%%%%%%%%%%%%%%%%%%%
%%%%%%%%%%%%%%%%%%%%%%%%%%%%%%%%%%%%%%
%%%%%%%%%%%%%%%%%%%%%%%%%%%%%%%%%%%%%%
%%%%%%%%%%%%%%%%%%%%%%%%%%%%%%%%%%%%%%

\begin{table}[tbp]
\centering
\setlength{\extrarowheight}{0pt}
\addtolength{\extrarowheight}{\aboverulesep}
\addtolength{\extrarowheight}{\belowrulesep}
\setlength{\aboverulesep}{0pt}
\setlength{\belowrulesep}{0pt}
\caption{The Quantum-Inspired Technology for Database and Information Retrieval (IR)}
\label{tab:quantumInspired}
\begin{tabular}{m{1.5em}<{\centering}m{7em}<{\centering}m{12em}<{\centering}m{15em}<{\centering}}
\hline
% \rowcolor{lightgray!20}
\rowcolor[gray]{.9}

\multicolumn{1}{c}{} & \textbf{Time} & \textbf{Works} & \textbf{Problems} \\
\toprule

\multirow{3}{*}{\rotatebox[origin=c]{90}{\textbf{Database}}} & 2013 & \textit{Roy et al.} \cite{DBLP:conf/cidr/0002KK13} & Resource Allocation \\

\cline{2-4} 
    & 2020 / 2021 &\textit{Yuan and Lu} \cite{yuan2020quantum,yuan2021quantum} &  Keyword Searches  \\ 

\cline{2-4} 
    & 2020 / 2021 & \textit{Sayed and Mhmed} \cite{mohsin2020dynamic,mohsin2021dynamic} &  Query Optimization  \\ 
\toprule

\multirow{4}{*}{\rotatebox[origin=c]{90}{\textbf{IR}}} & 2004 &  \textit{Rijsbergen} \cite{DBLP:books/daglib/0011947}  &  IR  \\

\cline{2-4} 
    & 2013  & \textit{Sordoni et al.} \cite{10.1145/2484028.2484098} &  IR (Quantum Language Model)  \\ 

\cline{2-4} 
    & 2010  & \textit{Wang et al.} \cite{DBLP:conf/aaaifs/WangSK10} &  Multimodal IR  \\ 

\cline{2-4} 
    & 2018 & \textit{Wang et al.} \cite{DBLP:conf/ecir/WangWH018} &  User Interactions in IR  \\ 
\hline

\bottomrule
\end{tabular}
\end{table}

% %A quantum-based database query scheme for privacy preservation in cloud environment 2019

% -------------------------------------------
% \cite{liu2019quantum}

% -------------------------------------------

%%%%%%%%%%%%%%%%%%%%%%%%%%%%%%%%%%%%%%
%%%%%%%%%%%%%%%%%%%%%%%%%%%%%%%%%%%%%%
%%%%%%%%%%%%%%%%%%%%%%%%%%%%%%%%%%%%%%
%%%%%%%%%%%%%%%%%%%%%%%%%%%%%%%%%%%%%%

%How the quantum-inspired framework supports keyword searches on multi-model databases 2020
%Quantum-Inspired Keyword Search on Multi-model Databases 2021

\textbf{\textit{Yuan and Lu}} \cite{yuan2020quantum,yuan2021quantum} use a quantum-inspired framework exploiting non-classical probabilities to support keyword searches on multi-model databases. This work represents relational data, JSON documents, and Graph data with the help of the quantum language model. Then it could bring the problem into vector spaces and describe events as subspaces. Next, this framework utilizes density matrices to encapsulate all the information over these subspaces. It uses those density matrices to measure the divergence between query and candidate answers to find top-\textit{k} the most relevant results. Considering the complexity of matrices computation, they utilize a vector form deduced from the probability formulation to describe data and density. Besides, this work uses spatial pattern mining technology to identify compounds and takes Principle Component Analysis (PCA) methods to reduce computation dimension for helping quantum-inspired keyword searches have better performance.

%%%%%%%%%%%%%%%%%%%%%%%%%%%%%%%%%%%%%%
%%%%%%%%%%%%%%%%%%%%%%%%%%%%%%%%%%%%%%
%%%%%%%%%%%%%%%%%%%%%%%%%%%%%%%%%%%%%%
%%%%%%%%%%%%%%%%%%%%%%%%%%%%%%%%%%%%%%

% Dynamic cost ant colony algorithm for optimize distributed database query 2020
% Dynamic cost ant colony algorithm to optimize query for distributed database based on quantum-inspired approach 2021

\textit{\textbf{Sayed and Mhmed}} \cite{mohsin2020dynamic,mohsin2021dynamic} propose a quantum-inspired ant colony algorithm aiming to identify the most efficient join order to minimize the overall cost of the query plan on distributed databases. To reach the goal, this work utilizes a sum of the I/O cost (calculated for all join processes) and data transfer cost (transferring entities/tables between database sites) as the total cost. For the search space, it is implemented as a graph $G = (N, E)$ where $N$ is the collection of nodes (entities/tables in the query specification) and $E$ is the set of edges (join relationship between corresponding tables). And the search methodology is the quantum-inspired ant colony algorithm where each entity is associated with a qubit, and the quantum partial negation gate is employed to update the entity’s qubit amplitudes. In detail, each ant randomly chooses an entity as the beginning point. Then, a partial negation quantum gate is used to determine the next entity. Next, the join cost is calculated, and this new entity will be next beginning of the following ant. Repeat this process till all entities are handled, and select the best journey cost from all the journeys for ants. After that, update the pheromone and repeat the whole steps until maximum iterations. Finally, identify the best trail for all ants, which is the solution.

\paragraph{\textbf{Discussion}} 

In this section, we have reviewed the works \cite{DBLP:conf/cidr/0002KK13,yuan2020quantum,yuan2021quantum,mohsin2020dynamic,mohsin2021dynamic} together. In these works, 
\cite{DBLP:conf/cidr/0002KK13} optimizes resource allocation in light of the notion of entanglement (queries and transactions), not computational way. \cite{yuan2020quantum,yuan2021quantum}  utilize quantum physics's probabilistic formalism for keyword searches on multi-model databases. Although it employs some technologies to reduce the complexity of algorithms, it still needs lots of effort and time to prepare and perform. \cite{mohsin2020dynamic,mohsin2021dynamic} focus on the query optimization. Instead of using the probabilistic mechanism in its processing, it uses the quantum partial negation gate to build its algorithms. In its model, performance improvement is reached, but at the expense of time.

%%%%%%%%%%%%%%%%%%%%%%%%%%%%%%%%%%%%%%
%%%%%%%%%%%%%%%%%%%%%%%%%%%%%%%%%%%%%%
%%%%%%%%%%%%%%%%%%%%%%%%%%%%%%%%%%%%%%
%%%%%%%%%%%%%%%%%%%%%%%%%%%%%%%%%%%%%%

% \subsection{Other Studies on Quantum-inspired algorithm}

% We can add some papers from quantum-inspired  information retrieval 
% See the following paper:
%https://arxiv.org/pdf/2007.04357.pdf

% The quantum-inspired algorithm is a growing area. A wide array of work in different aspects of disciplines has been done and produced promising results. In addition to the above quantum-inspired works on databases, here are some papers in the field of information retrieval. 

\subsection{Quantum Inspired Technology for Information Retrieval (IR)}

The quantum-inspired algorithm is a growing area. A wide array of work in different aspects of disciplines has been done and produced promising results. In addition to the above works related to databases, here are some papers in the field of information retrieval (IR) \cite{10.1145/3402179}.

%%%%%%%%%%%%%%%%%%%%%%%%%%%%%%%%%%%%%%
%%%%%%%%%%%%%%%%%%%%%%%%%%%%%%%%%%%%%%
%%%%%%%%%%%%%%%%%%%%%%%%%%%%%%%%%%%%%%
%%%%%%%%%%%%%%%%%%%%%%%%%%%%%%%%%%%%%%
 
%The geometry of information retrieval 2004
\textbf{\textit{Rijsbergen}} \cite{DBLP:books/daglib/0011947} offers the original idea regarding using the mathematical framework of quantum theory in information retrieval. He aims to investigate a formal description of user interactions and the abstraction of the concept of a document in IR. 
To make the same event have multiple representations, he uses the Hilbert space representation in the Quantum framework to abstract and contextualize information objects like documents and queries. Besides, he also demonstrates how the Quantum-like formulation can be applied to perform existing IR tasks, including coordination level matching, feedback, clustering, and more.

% Modeling Term Dependencies with Quantum Language Models for IR 2013
\textbf{\textit{Sordoni et al.}} \cite{10.1145/2484028.2484098} present the Quantum Language Model (QLM), combining the Vector Space Model and Probabilistic Language Model of classical IR through the Hilbert space formalism. In QLM, a document is described as a sequence of projectors, where a projector means a single term or compound term from the document, and a compound term is defined as superposition events through the quantum generalization of probabilities. Then a density matrix $\rho$ can be obtained based on the projectors of a document. And a language model is essentially a density matrix $\rho$. Similarly, a language model for a query $\rho_q$ can be obtained, too. Finally, with the help of a scoring function (e.g., quantum relative entropy), the relevance of a document for a query can be calculated.

% A Quantum Theory Inspired Image Retrieval Framework   2010

\textbf{\textit{Wang et al.}} \cite{DBLP:conf/aaaifs/WangSK10} focus on multimodal information retrieval, where a document may include textual and visual content features. They exploit the tensor product of Hilbert spaces to combine textual and visual features of an image like a non-separable composite system. Besides, they propose two statistical strategies to identify the correlation between text and visual features and embed it into the density matrix of the composite system: \ding{182} Maximum Feature Likehood that associates the text with the maximal likely visual feature; \ding{183} Feature and Word Mutual Information Matrix that describes the dependency between the visual and textual features.

% Modeling Relevance Judgement Inspired by Quantum Weak Measurement 2018

\textbf{\textit{Wang et al.}} \cite{DBLP:conf/ecir/WangWH018} perform a user experiment to study the variance of relevance judgment and get the conclusion that the quantum weak measurement is more appropriate to model relevance judgment compared to the standard quantum measurement. This is because the weak measurement has a larger variance, which the authors believe could help explain complex cognitive phenomena, i.e., a user's information need is dynamic. Besides, to model the user's dynamic evolving information needs, they also propose a weak measurement-based session search model.

% Although those papers report an application of the proposed representation of entities like documents, queries, etc., in information retrieval, it would be interesting to use it to construct information representation in databases.

\paragraph{\textbf{Discussion}} The application of quantum-inspired technology in information retrieval (IR) has been explored by several researchers, offering innovative perspectives on user interactions, document abstraction, and multimodal information retrieval. Rijsbergen \cite{DBLP:books/daglib/0011947} pioneers the use of the mathematical framework of quantum theory in IR, leveraging the Hilbert space representation to abstract and contextualize information objects, such as documents and queries. Building on this foundation, Sordoni et al. \cite{10.1145/2484028.2484098} introduced the Quantum Language Model, which combines the Vector Space Model and Probabilistic Language Model of classical IR through the Hilbert space formalism. This model represents documents as sequences of projectors and utilizes density matrices to describe language models for both documents and queries. Wang et al. \cite{DBLP:conf/aaaifs/WangSK10} extend the quantum framework to multimodal information retrieval, incorporating textual and visual content features into consideration. They utilize the tensor product of Hilbert spaces to represent non-separable composite systems and proposed statistical strategies to identify correlations between text and visual features, embedding these relationships into a density matrix. Furthermore, Wang et al. \cite{DBLP:conf/ecir/WangWH018} conduct a user experiment to investigate the variance of relevance judgment, concluding that quantum weak measurement is a more suitable model for relevance judgment than standard quantum measurement. The integration of quantum theory in IR offers a promising direction for future research. By leveraging quantum principles, researchers can develop advanced models and algorithms that enhance the efficiency and capability of information retrieval systems, ultimately providing users with more effective and personalized search experiences. Although those papers report applications of the proposed representation of entities (e.g., documents and queries) and interaction in information retrieval, it would be interesting to use them to construct information representation and interaction in databases.

%Although the paper reports an application of the proposed language model in speech recognition, it would be interesting to use it to construct document and query language models.

% The improvement in performance ranged, in some cases, between 77% and 99%, but at the expense of time. The tensor product used in QIACO affects the search time and needs more time to reach the best QEP than the classical ACO. Additionally, the results imply that, although the classical ACO is used successfully with simple joins that have a small number of join entities, QIACO can also be used with simple and complex queries with numerous join entities. 

% To solve the problems in traditional methods, the quantum-inspired ant colony (QIACO) paradigm is used in try to reach the optimum query optimization. Here, quantum-inspired is employed to change the seeking procedure used by the classical ant colony algorithm to move from one node to another. Instead of using the probabilistic mechanism while building the ant solution, our algorithm will use the quantum partial negation gate, controlled by pheromone values, to control the ant movement. 

\section{Open Challenges and Future Work}\label{sec:open-challenges_future-work}

% Despite the potential advantages of quantum computing in database management systems, several open challenges need to be addressed to facilitate the practical implementation of quantum databases:
% There are several challenges and future work that need to be addressed before applying quantum computing to multi-model databases:
% Lastly, we discuss the research challenges involved in making the quantum database abstraction a real-world tool for developers (Section 6). While these challenges are significant, we believe that they are possible to resolve and that working on them will yield new insights into multiple aspects of database systems, well beyond resource allocation applications.

Implementing databases with quantum computing presents several challenges that need to be addressed. While these challenges may be difficult, we believe they are solvable, and resolving them will provide new insights into various aspects of database systems. 

\begin{itemize}
    \item \textbf{Hardware Research:} Current quantum computing hardware has limited qubit counts, coherence times, error rates, and I/O bandwidth, which restrict the applicability of quantum computing to real-world databases. We should find ways to improve the hardware capabilities to enable large-scale, practical quantum computing-related databases.

    \item \textbf{Quantum Data Models and Query Languages:} Developing quantum data models and query languages tailored for quantum computing is crucial to enable seamless integration with existing database systems. Researchers need to devise standardized models and languages that can express simple or even complex database operations and take advantage of quantum computing capabilities. Besides, exploring quantum analogs of classical concepts like integrity used in database systems is also challenging \cite{arxiv.0705.4303}. 
    
    % Of course, we could assign new meanings to some old concepts. For example, quantum integrity could be framed as the fidelity of quantum information throughout its quantum state evolution, incorporating both quantum superposition and entanglement. 

    \item \textbf{Scalability and Robustness:} Quantum algorithms and data structures must be designed to be scalable and robust in the presence of noise and errors. Future research should focus on error-tolerant algorithms and techniques to mitigate the effects of hardware imperfections on quantum computing-related database performance.

    \item \textbf{Interoperability and Hybrid Systems:} As quantum computing matures, there will be a need for hybrid systems that can leverage both classical and quantum resources efficiently. Researchers should explore methods to ensure smooth interoperability between classical and quantum components, allowing for a gradual transition towards fully quantum databases (storing and processing data on quantum machines).

    \item \textbf{Leveraging Quantum Computing-Inspired Methods in Database:} Quantum-inspired computing incorporates certain principles and algorithms inspired by quantum mechanics into classical computing systems, offering a glimpse into the power of quantum computing without requiring complex hardware, which also offers a pathway to improve the performance and capabilities of database systems. However, except for the previous mention (Resource Allocation, Keyword Searches, and Query Optimization), it still needs creative thinking on how to use this kind of technology to improve the performance of database systems.
    
\end{itemize}
% Educating the Workforce: As quantum databases become more prevalent, there will be a growing demand for professionals skilled in quantum computing and database management. It is essential to invest in education and training programs to prepare the workforce for the upcoming quantum revolution in the database industry.

We have identified several open problems for quantum databases in the preceding discussion. These challenges present opportunities for future research to advance the practicality and real-world applicability of quantum database abstraction.

% In conclusion, while quantum computing has many potential applications in multi-model databases, more work is needed before it can be applied in a practical and scalable way. Researchers need to address the challenges listed above to make quantum database abstraction a real-world tool for developers.

\begin{itemize}
    \item \textbf{Hardware Accelerator:} Similar to Sch{\"{o}}nberger et al. \cite{sigmod2023}, developers could design quantum computing as a specialized hardware component of classical databases to accelerate data processing. In this way, quantum computing could help improve classical database performance (i.e., developing \textit{quantum-accelerated databases}). Additionally, it is worth studying how to translate quantum algorithms into circuits, which could work well (e.g., reducing the error inherited by the different circuits’ hardware realization) on the corresponding quantum hardware \cite{DBLP:conf/adbis/JoczikK20}.

    \item \textbf{Quantum Vector Database:} It is conceivable to hypothesize the existence of quantum databases dedicated to vector data storage. Vector representations are ubiquitously employed to numerically transcribe multi-modal data, which spans a multitude of formats including, but not limited to, video, image, auditory, and textual data. Quantum computing confers the capability to execute data processing tasks, conventionally time-consuming on classical database systems, concurrently on a quantum computer. By integrating the principles of quantum computing with the concept of vector databases, we foresee potential advancements in domains such as similarity search and data retrieval. However, based on this model, how to extract useful information from a quantum computer in a superposition or how to effectively and efficiently remove some items from a superposition in a single step is still waiting for researchers to solve. Besides, designing a quantum algorithm that can match queries with an index is urgently needed.

    % The idea of a quantum vector database leverages the principles of quantum superposition and entanglement to store and process vector data more efficiently. 

    % Quantum computers, with their inherent parallelism and exponentially large Hilbert spaces, can potentially provide superior solutions for storing vector data and processing it for similarity searches and retrieval tasks. 

    %We could envision quantum databases for storing vector data. Vectors are commonly used to encode multi-modal data, encompassing diverse forms such as video, image, music, and text, into numerical representations. Then, quantum Computing gives the ability to perform data processing, which usually takes a long processing time on a classical database system, in a simultaneous way on a quantum computer. By combining the principles of quantum computing with the concept of vector databases, we anticipate advancements in the fields of similarity search and data retrieval.

    \item \textbf{Quantum Distributed Database:} The concept of a distributed quantum database may be an exciting development. With the potential for near-instantaneous transfer of quantum database files in a superposition state, this new design could revolutionize the way we think about database processing and manipulation and offer a more efficient way of handling large-scale databases. Besides, quantum communication technology based on quantum entanglement and quantum teleportation could be used to realize the secure transmission of information.

    \item \textbf{Quantum AI for DB:} Recently, many researchers in the field of quantum computing have put a lot of energy into quantum artificial intelligence \cite{DBLP:conf/sigmod/WinkerGUYLFM23, massoli2022leap}, expecting to accelerate the training. Then, it may be a nice attempt to leverage proposed quantum AI techniques to improve database management systems. Quantum AI can be employed to optimize database operations, such as query processing, data analytics, and pattern recognition. For example, quantum machine learning algorithms can be utilized to identify correlations and trends in large-scale databases, enabling efficient indexing and data summarization. Furthermore, quantum AI can aid in developing intelligent caching and prefetching strategies, which can significantly improve the performance of database systems. By integrating quantum AI into database management systems, researchers can unlock new possibilities in data processing and analytics, ultimately enabling more capable and responsive database systems.

    \item \textbf{Quantumn Database for AI:} These days, some people work on ``databases (DB) for artificial intelligence (AI)'', i.e., making databases service machine learning models. Similarly, how to make quantum databases (QDB) service (quantum) artificial intelligence could be a potential direction. This is because as AI models become increasingly complex and data-intensive, the need for efficient storage and retrieval of vast amounts of data becomes paramount. quantum databases can provide an exponential speed-up in data access and processing, allowing AI systems to train and make predictions faster and more accurately. Moreover, quantum databases can facilitate secure storage and sharing of sensitive data, such as medical records or financial transactions, which can be invaluable for AI applications in healthcare, finance, and other domains with strict data privacy requirements. Envisioning further work, researchers could focus on developing efficient interfaces between quantum databases and AI models, enabling seamless data exchange and fostering the growth of quantum-enhanced AI applications.

    % Recently, people have explored databases (DB) for artificial intelligence (AI), using databases to store data for machine learning models. Quantum computing could be used to accelerate the training of machine learning models by performing optimization tasks more efficiently. Hence, how to integrate quantum computing into databases to help train may be an interesting research point.

    \item \textbf{Quantum Graph Database:} Graph databases are used to store and process graph structures, such as social networks, web pages, and molecular structures. Most researchers are studying combinatorial optimization problems (e.g., Hamiltonian Cycle Problem \cite{10.1145/3290688.3290703}, Traveling Salesman Problem \cite{tsp-grover}, Sub-graph Isomorphism Problem \cite{10.1145/3569095}, etc.) with quantum computing, which is helpful in building graph processing mechanisms in quantum graph databases.

    \item \textbf{Security:} Databases often contain sensitive data that needs to be encrypted to ensure data privacy. We could use quantum computing technologies to develop more secure yet efficient encryption strategies that are resistant to attacks by quantum computers. However, while quantum encryption methods can provide provable security against private search, they may be vulnerable to quantum attacks as quantum computers become more powerful. Additionally, integrating quantum encryption methods with existing database systems can be challenging and may require significant changes to the system architecture.

    \item \textbf{Query Optimization:} After reviewing the literature, there are a few works on query optimization (i.e., Multiple Query Optimization and Joining Order). In fact, query optimization is a critical task in databases, which features a broad range of complex optimization problems with large solution spaces. The advantage of quantum computing is that it could be used to perform complex optimization tasks that are difficult to solve using classical computing. Hence, using quantum computing could be exciting research for other query optimization problems such as finding proper query plans, evaluating the cost of query plans, or designing indices.

    \item \textbf{Quantum Computing-Inspired Databases:} As research and development in the field of quantum-inspired computing continue to progress, we can expect even more innovative solutions that push the boundaries of database performance, further revolutionizing the way we process, analyze, and secure data. The potential directions we could set out to do are: Quantum-inspired algorithms, such as quantum-inspired heuristics \cite{8943210}, can be employed to more effectively explore a vast search space of possible query plans, leading to optimal or near-optimal solutions;
    Quantum-inspired indexing techniques, such as quantum-inspired hashing \cite{9626574}, can be employed to enhance the efficiency of traditional indexing structures;
    Quantum-inspired machine learning algorithms, such as quantum-inspired support vector machine \cite{9451546} can be employed to enhance the performance of data analysis tasks, such as clustering, classification, and anomaly detection.
    
    % Quantum-inspired encryption algorithms, such as lattice-based cryptography, can be employed to provide stronger encryption techniques that are resistant to attacks by quantum computers.

\end{itemize}

\section{Conclusion}\label{sec:conclusion}

Quantum computing has the potential to impact various aspects of database management systems. The quantum databases born out of this fact hold the promise to revolutionize the field of database management by leveraging the unique capabilities of quantum computing. Although quantum databases are still in the early stages of development, and there are still many challenges to be overcome in determining how quantum computing can be applied to databases in practical and scalable ways, the ongoing research efforts and growing interest from academics indicate a bright future for quantum databases. Considering the current lacking a comprehensive review of the literature in this emerging field and the individual works are largely segmented, we aim to, in this paper, provide a detailed overview of the history, current situation, problems, and ongoing efforts in the development of quantum databases, review the literature systematically to provide a clear picture of the landscape and a road-map for the future, shedding light on the opportunities and challenges that lie ahead in this exciting domain. We aspire that this comprehensive review will serve as a catalyst, inspiring further contributions from researchers, and thereby propelling advancements in this exciting field.

\bibliographystyle{abbrv}
\bibliography{main}

\end{document}